\documentclass[%
 reprint,
nofootinbib,
 amsmath,amssymb,
 aps,
floatfix,
]{revtex4-1}

\usepackage{graphicx}
\usepackage{dcolumn}
\usepackage{bm}

\begin{document}

\preprint{APS/123-QED}

\title{On the group theoretic structure of a class of quantum dialogue protocols}

\author{Chitra Shukla}
\author{Vivek Kothari}
 \affiliation{%
 Jaypee Institute of Information Technology, A-10, Sector-62,
Noida, UP-201307, India.}
%

\author{Anindita Banerjee}
\affiliation{
Department of Physics and Center for Astroparticle Physics
and Space Science, Bose Institute, Block EN, Sector V, Kolkata 700091,
India.
}%

\author{Anirban Pathak}
\email{anirban.pathak@upol.cz}
\affiliation{%
Jaypee Institute of Information Technology, A-10, Sector-62,
Noida, UP-201307, India.
}%
\affiliation{%
 RCPTM, Joint Laboratory of Optics of Palacky University and
Institute of Physics of Academy of Science of the Czech Republic,
Faculty of Science, Palacky University, 17. listopadu 12, 77146 Olomouc,
Czech Republic.
}%

\date{\today}

\begin{abstract}
Intrinsic symmetry of the existing protocols of quantum dialogue are
explored. It is shown that if we have a set of mutually orthogonal
$n$-qubit states {\normalsize $\{|\phi_{0}\rangle,|\phi_{1}\rangle,....,|\phi_{i}\rangle,...,|\phi_{2^{n}-1}\rangle\}$
and a set of $m-qubit$ ($m\leq n$) unitary operators $\{U_{0},U_{1},U_{2},...,U_{2^{n}-1}\}:U_{i}|\phi_{0}\rangle=|\phi_{i}\rangle$
and $\{U_{0},U_{1},U_{2},...,U_{2^{n}-1}\}$ forms a group under multiplication
then it would be sufficient to construct a quantum dialogue protocol
using this set of quantum states and this group of unitary operators}.
The sufficiency condition is used to provide a generalized protocol
of quantum dialogue. Further the basic concepts of group theory and
quantum mechanics are used here to systematically generate several
examples of possible groups of unitary operators that may be used
for implementation of quantum dialogue. A large number of examples
of quantum states that may be used to implement the generalized quantum
dialogue protocol using these groups of unitary operators are also
obtained. For example, it is shown that $GHZ$ state, $GHZ$-\emph{like}
state, $W$ state, 4 and 5 qubit Cluster states, $\Omega$ state,
Brown state, $Q_{4}$ state and $Q_{5}$ state can be used for implementation
of quantum dialogue protocol. The security and efficiency of the proposed
protocol is appropriately analyzed. It is also shown that if a group of unitary operators and a set of mutually orthogonal states are found to be suitable for quantum dialogue then they can be used to provide solutions of socialist millionaire problem.

\end{abstract}

\pacs{Valid PACS appear here}

\keywords{Suggested keywords}
\maketitle

\section{\label{Introduction}Introduction}

First protocol of unconditionally secure quantum key distribution
(QKD) was proposed by Bennett and Brassard in 1984 \cite{bb84}. In
a QKD protocol two remote legitimate users (Alice and Bob) can establish
an unconditionally secure key by using quantum resources i.e. by the
transmission of qubits. The protocol of Bennett and Brassard which
is popularly known as BB-84 protocol, draw considerable attention
of the cryptographic community since the unconditional security of
key obtained in this protocol is not achievable in classical cryptography.
Naturally, since 1984 several new protocols for different cryptographic
tasks have been proposed. While most of the initial works on quantum
cryptography \cite{bb84,ekert,b92} were concentrated around QKD,
eventually quantum-states were applied to other `post-coldwar' cryptographic
tasks. For example, in 1995, Goldenberg and Vaidman \cite{vaidman-goldenberg} proposed a protocol of quantum secure direct communication (QSDC), then in 1999, a protocol for quantum secret sharing
(QSS) was proposed by Hillery \cite{Hillery}. In the same year, Shimizu
and Imoto \cite{Imoto} proposed a protocol for deterministic secure quantum
communication (DSQC) using entangled photon pairs. In the Goldenberg-Vaidman protocol and in Shimzu-Imoto protocol Alice
can communicate a message to Bob directly with unconditional security.
Other important protocols for QSDC  were
later proposed \cite{ping-pong,lm05}.  In a DSQC protocol,
the receiver can read out the secret message only after the transmission
of at least one bit of additional classical information for each qubit.
A  QSDC  protocol  does not
require exchange of classical information.  Since the pioneering works of Goldenberg and Vaidman \cite{vaidman-goldenberg} and Shimizu and Imoto \cite{Imoto} several
protocols of DSQC and QSDC are proposed (\cite{review} and references
there in). But in all these QSDC and DSQC protocols, the meaningful
information (secret message) travels only from Alice to Bob%
\footnote{The protocol may be a two way protocol like Ping-Pong protocol \cite{ping-pong}
or LM-05 \cite{lm05} protocol but the meaningful information (message)
is transmitted from Alice to Bob only. Thus the flow of information
is unidirectional (one way) only.%
}. In other words in these protocols, Alice and Bob can not simultaneously
transmit their different secret messages to each other (dialogue)
and consequently advent of these protocols naturally leads to a question:
Is it possible to extend these protocols for bidirectional quantum
communication in which both Alice and Bob will be able to communicate
(with unconditional security) using the same quantum channel. Such
bidirectional protocols are quantum dialogue protocols where information
can flow along two directions (i.e. from Alice to Bob and from Bob
to Alice). Such protocols are actually an essential requirement of
our everyday communication problems. This can be visualized more clearly
if we consider the analogy of a telephone. The possibility of extending
the DSQC and QSDC protocols and the absolute need of bidirectional
quantum communication motivated the quantum communication community
to investigate the possibility of designing of quantum dialogue protocols.
First protocol of quantum dialogue was proposed by Ba An \cite{ba an}
using Bell states in 2004. Eventually it was found that the protocol
is not secure under intercept-resend attack \cite{Man}. But a minor
modification of the protocol can make it unconditionally secure and
such modified protocol was first proposed by Man, Zhang and Li \cite{Man}.
Later on Xia \emph{et al.} proposed a protocol of quantum dialogue
using $GHZ$ states \cite{xia} and Dong \emph{et al.} proposed a
protocol of quantum dialogue using tripartite $W$ states \cite{dong-w}.
But in essence all these protocols are same. Here we will refer to
all these protocols as Ba An type of protocol and provide a general
structure to them. In recent past several protocols of quantum dialogue
have been proposed using i) dense coding \cite{Man,xia,quantum telephon1},
ii) entanglement swapping \cite{gao-swapping}, iii) single photon
\cite{Naseri}, iv) auxiliary particles \cite{shi-auxilary} etc.
These protocols are referred as bidirectional quantum communication
protocols \cite{shi-auxilary}, quantum telephone \cite{quantum telephon1,Y Sun improve telephone},
quantum dialogue \cite{ba an,Naseri}, quantum conversation \cite{sakshi-panigrahi-epl}
etc. These are different names used for equivalent protocols. Here
we will refer all of them as quantum dialogue and provide a generalized
structure to the Ba An type of quantum dialogue protocols and will
use the generalized structure to obtain several examples of quantum
systems where quantum dialogue is possible. Before we describe those
specific quantum system it is important to understand that in quantum
dialogue the communication between Alice and Bob is simultaneous.
The simultaneity implies that quantum channel (i.e. the quantum states
on which the classical information of Alice and Bob are encoded) must
simultaneously contain the information encoded by both parties. This
particular point distinguishes quantum dialogue protocol from the
QSDC and DSQC protocols. Otherwise, Alice and Bob can always communicate
with each other by using DSQC/QSDC in two steps or
by using two different quantum channels (i.e. by using a DSQC/QSDC
scheme from Alice to Bob and another from Bob to Alice) but as the
secret information of Alice and Bob are not simultaneously encoded
in the same quantum channel, this is not quantum dialogue. This important
and distinguishing feature of quantum dialogue is often overlooked
by authors. For example, Jain \emph{et al.}'s \cite{sakshi-panigrahi-epl} protocol
is essentially two QSDC. Clearly, their protocol is not a protocol
of quantum dialogue as Bob knows the encoded information of Alice
even before he encodes his own information.

The remaining part of the paper is organized as follows: In Section
\ref{protocol}, we have briefly described the Ba An protocol and have explored
its intrinsic symmetry. We have observed that information splitting
is in the core of these protocols%
\footnote{Information splitting plays the central role in every secure\\ quantum
communication protocol.%
}. In Section \ref{sufficient condition} we have provided a sufficient condition for construction
of quantum dialogue protocol and have shown that the operators used
for encoding of information in a quantum dialogue protocol should
form a group. In Section \ref{general protocol} we have provided a generalized protocol
of the quantum dialogue. To implement the protocol we require a set
of unitary operators that form a group under multiplication and a
set of mutually orthogonal states on which the information are to
be encoded by these group of unitary operators. A systematic procedure
for construction of such groups and specific examples of states that
can be used to implement the generalized protocol of quantum dialogue
are provided in Section \ref{group construction}. It is shown that $GHZ$ state, $GHZ$-\emph{like}
state, $W$ state, Cluster state, $\Omega$ state, $Q_{4}$ state
and $Q_{5}$ state can be used for implementation of quantum dialogue
protocol. Finally Section \ref{conclusion} is dedicated for conclusion where we have concluded our discussion about the group theoretic structure of quantum dialogue protocols and  have shown that if a group of unitary operators and a set of mutually orthogonal states are found to be suitable for quantum dialogue then they can be used to provide solutions of socialist millionaire problem too.

\section{The Ba An protocol and its intrinsic symmetry}\label{protocol}

Let us first describe Ba An's original scheme of quantum dialogue.
This simple scheme works in following steps:
\begin{enumerate}
\item Bob prepares large number of copies of a Bell state $|\phi^{+}\rangle=\frac{|01\rangle+|10\rangle}{\sqrt{2}}$.
He keeps the first photon of each qubit with himself as home photon
and encodes her secret message $00,01,10$ and $11$ by applying unitary
operations $U_{0},U_{1},U_{2}$ and $U_{3}$ respectively on the second
qubit. Without loss of generality we may assume that $U_{0}=I,\, U_{1}=\sigma_{x}=X,\, U_{2}=i\sigma_{y}=iY$
and $U_{3}=\sigma_{z}=Z$ where $\sigma_{i}$ are Pauli matrices.
\item Bob then sends the second qubit (travel qubit) to Alice and confirms
that Alice has received a qubit.
\item Alice encodes her secret message by using the same set of encoding
operations as was used by Bob and sends back the travel qubit to Bob.
After receiving the encoded travel qubit Bob measures it in Bell basis.
\item Alice announces whether it was run in message mode (MM) or in control
mode (CM). In MM, Bob decodes Alice\textquoteright{}s bits and announces
his Bell basis measurement result. Alice uses that result to decode
Bob's bits. In CM, Alice reveals her encoding value to Bob to check
the security of their dialogue.
\end{enumerate}
It is easy to recognize that this is a modification of Ping-Pong protocol
and the operations used for encoding are the operators usually used
for dense coding and the protocol starts with an initial state $|\psi\rangle_{initial}=|\phi^{+}\rangle$.
Now after Step 1, $|\phi^{+}\rangle$ is mapped to one of the Bell
state $|\psi\rangle_{intermediate}=U_{B}|\psi\rangle_{initial}=U_{B}|\phi^{+}\rangle$
depending upon the secret message of Bob which is encoded by unitary
operation $U_{B}$ (to be precise, we may say that the state at this
time is one of the Bell state $I|\phi^{+}\rangle=|\phi^{+}\rangle,\, X|\phi^{+}\rangle=|\psi^{+}\rangle,\, iY|\phi^{+}\rangle=|\psi^{-}\rangle,\, Z|\phi^{+}\rangle=|\phi^{-}\rangle$).
Thus in the second step, second qubit of one of the Bell states (one
of the mutually orthogonal states) is communicated to Alice via the
quantum channel. At this stage neither Alice nor Eve can know what
information is sent by Bob as they have access to only one qubit of
the entangled pair. Now in Step 3 Alice encodes her message using
the same set of unitary operations and Alice's encoding will map the state
into another Bell state ($|\psi\rangle_{final}=U_{A}|\psi\rangle_{intermediate}=U_{A}U_{B}|\psi\rangle_{initial}=U_{A}U_{B}|\phi^{+}\rangle$.
Now here information splitting is done in an excellent way. Alice,
Bob and Eve, all know $|\psi\rangle_{initial}$ and $|\psi\rangle_{final}$
states. But in addition, Alice and Bob knows the unitary operators
used by them for encoding. Availability of these additional information
allows them to decode each other's information and lack of this information
makes it impossible for Eve to decode the information encoded by Alice
and Bob. To make it more clear, assume that $|\psi\rangle_{final}=|\phi^{+}\rangle$
thus $U_{A}U_{B}=I,$ this is possible in 4 different ways: $U_{A}=U_{B}=I,\, U_{A}=U_{B}=X,\, U_{A}=U_{B}=iY,\, U_{A}=U_{B}=Z$.
Thus from the initial state and final state Alice and Bob can come to
know the encoding of each other but for Eve all encodings are possible.
She just obtains a correlation between the encoding of Alice and that
of Bob. In this particular example, Eve knows that Alice and Bob have
encoded the same message (same classical bits in this particular example),
but that does not reveal the encoding of Alice and Bob. Since in this
quantum dialogue protocol secure classical information (4 bits of
classical information in this case as 2 bits are send from Alice to
Bob and 2 bits are send from Bob to Alice) that is communicated using
the quantum channel is more than the dense coding capacity of the
quantum channel, so it is obvious that some correlation between Alice's
encoding and Bob's encoding will be obtained by Eve. This is recently
pointed out by Tang and Cai \cite{ijqi-tang}. But the information
splitting is done in such a way that the correlation does not directly
leak the encoding. To be precise, in the above example, even after
knowing the correlation (i.e. both Alice and Bob have encoded the same
classical message) Eve will not be able to develop any procedure to
obtain the encoding of Alice. For Eve all the encoding of Alice are
equally probable. So she has to guess randomly. If her guess is correct
(whose probability is $1/4)$ only then she will correctly obtain
Bob's encoded information. Now the probability of Eve's success can
be reduced by three means: 1) Using multipartite entangled states.
For example, if we use $2^{5}$ unitary operations and 5 qubit Brown
state \cite{Brown} to implement
quantum dialogue protocol then the success probability of Eve would
be $\frac{1}{32}$ only. This is so because after obtaining the correlation
Eve has to guess among 32 equally probable alternatives. This point
would be more clear when we will describe the generalized protocol.
2) By encoding lesser amount of information in the quantum channel
(compared to the dense coding capacity of the quantum channel). In
that case it is obvious that Eve\textquoteright{}s mean information
gain on Alice and Bob\textquoteright{}s bits would reduce \cite{ijqi-tang}
and 3) by using both 1 and 2 above. Here it is important to note that
the existence of a classical correlation between the encoding of Alice
and Bob is an intrinsic problem to quantum dialogue protocols but
there are strategies (as mentioned above) to circumvent that problem
and essentially the security of all Ba An type of quantum dialogue
protocols arises from the above described process of information splitting.

The protocol of Ba An appears quite satisfactory up to this point.
But there is a problem, Eavesdropping check is done at the last stage
and the protocol is not safe under intercept-resend attack. This was
first pointed out by Man, Zhang and Yong \cite{Man}. The idea behind
the attack is simple: Eve intercepts the travel photon, keeps it with
herself and prepares a fake entangled pair in $|\phi^{+}\rangle.$
She keeps the home (first) photon of this fake entangled states and
sends the second one to Alice. As Alice can not distinguish it from
the qubit sent by Bob, she will encode her message. On the return
path Eve will intercept her fake travel photon and do a Bell measurement
on the fake entangled pair to obtain the message encoded by Alice.
Once Eve knows the unitary operation used by Alice she would apply
the same unitary operation on the actual travel photon and send it
back to Bob. After Bob announces publicly his Bell basis measurement
result, Eve can deduce Bob\textquoteright{}s bits. Thus the protocol
of Ba An is not secure under this intercept-resend attack. In order
to make it secure we have to change the strategy for Eavesdropping
checking. There exist two simple and equivalent strategies which can
be used to make the Ba An protocol secure: 1) From Bob to Alice communication,
Bob keeps some qubits as verification qubits and after confirming
that Alice has received the qubits, he announces the position of verification
qubits. Now verification qubits are measured by Alice randomly in
$\{|0\rangle,|1\rangle\}$ and $\{|+\rangle,|-\rangle\}$ basis. After
the measurement Alice announces the measurement outcome and the basis
used, then Bob measures the corresponding qubits in the same basis.
Looking at the correlation of the outcomes, Alice and Bob would be
able to determine the presence of Eve. 2) When Bob sends a sequence
of travel photons then he can insert equal number of decoy photons
randomly in the sequence of the travel photons. These decoy photons
are prepared randomly in $\{|0\rangle,|1\rangle,|+\rangle,|-\rangle\}$
states. After confirming that Alice has received all the qubits, Bob
announces the position of the decoy photons. After Bob's announcement
Alice randomly measures decoy qubits in $\{|0\rangle,|1\rangle\}$
and $\{|+\rangle,|-\rangle\}$ basis. After the measurement, Alice
announces the measurement outcomes and the basis used. If she has
chosen the correct basis (the same basis in which decoy state is prepared)
the measurement outcomes will be in perfect correlation with the decoy
states prepared by Bob. In any of the above two strategies, if Eve
is detected then Alice does not do her encoding operation and the
protocol is truncated. On the other hand, the absence of Eve in the
communication channel from Bob to Alice would essentially mean that
no fake photon sequence has been sent to Alice and that will make
the protocol secure against intercept-resend attack. The first strategy
was used by Man, Zhang and Li \cite{Man}. We will use the second
(decoy photon) strategy. So far we have seen that a clever choice
of information splitting plays the key role in the construction of
quantum dialogue protocols. To construct a generalized protocol of
quantum dialogue we need a deeper understanding of this information
splitting process. To be precise, a deeper understanding of the information
splitting process would help us to obtain a more general and sufficient
condition for construction of quantum dialogue protocol. Then just
by using standard tricks of Eavesdropping checking and the sufficient
conditions we will construct a generalized protocol of quantum dialogue.

\section{Sufficient condition for construction of quantum dialogue protocol}\label{sufficient condition}

Before we generalize the protocol we need to visualize certain intrinsic
symmetries and requirements of the Ba An type of protocols. In the
following discussion $|\psi\rangle_{initial}$ is an $n$-qubit state
in general, $|\psi\rangle_{intermediate}$ is the $n-qubit$ state
after encoding operation of Bob and $|\psi\rangle_{final}$ is an
$n-$qubit state after encoding of Alice. Thus they are not limited
to Bell states. Now, we can note that after encoding operations of
Bob, the initial state $|\psi\rangle_{initial}$ must be mapped to
a mutually orthogonal set of intermediate states $(|\psi\rangle_{intermediate})$.
Because only in that case, Alice would be able to deterministically
discriminate the intermediate states prepared by Bob and thus will
be able to decode the encoded message. To be precise, if we have $k$
unitary operations $U_{0},U_{1},U_{2},...,U_{k-1}$ and Bob encodes $i^{th}$
message by applying the unitary operator $U_{Bi}$, then after the
encoding operation the initial states are mapped to $U_{Bi}|\psi\rangle_{initial}=|\psi_{i}\rangle_{intermediate}:_{intermediate}\langle\psi_{i}|\psi_{j}\rangle_{intermediate}=\delta_{i.j}$.
Following the same logic, Alice's encoding operation $U_{Ai}$ should
also map the intermediate states to a mutually orthogonal set of final
state, i.e. $U_{Ai}|\psi\rangle_{intermediate}=|\psi_{i}\rangle_{final}:_{final}\langle\psi_{i}|\psi_{j}\rangle_{final}=\delta_{i},_{j}$.
Now the Hilbert space of $n-qubit$ states is $C^{2^{n}}.$ Therefore,
we can have at most $2^{n}$ mutually orthogonal states in that space.
Let us denote these states as $\left\{ |\phi_{0}\rangle,|\phi_{1}\rangle,....,|\phi_{i}\rangle,...,|\phi_{2^{n}-1}\rangle\right\} $.
It is easy to recognize that these states are nothing but the elements
of a mutually orthogonal basis set in $C^{2^{n}}.$ To make the remaining
discussion convenient, without loss of generality, we may assume that
at the beginning of the generalized protocol, Bob prepares a large
number of copies of the state $|\phi_{0}\rangle$. Now to encode an
arbitrary $n-bit$ classical message we would require $2^{n}$ unitary
operators. In principle one can chose to work with less number of
encoding operations but that would not have any effect other than
reducing the communication efficiency. So for an efficient protocol
we need $2^{n}$ number of $m-qubit$ unitary operations $\{U_{0},U_{1},U_{2},...,U_{2^{n}-1}\}$
where $m\leq n.$ As the operators are required to map the initial
state $|\phi_{0}\rangle$ into one of the state vector of the mutually
orthogonal set $\{|\phi_{0}\rangle,|\phi_{1}\rangle,....,|\phi_{i}\rangle,...,|\phi_{2^{n}-1}\rangle\}.$
Without loss of generality we can assume that $U_{i}|\phi_{0}\rangle=|\phi_{i}\rangle.$
If $m<n$ then we have dense coding, and if $m=\frac{n}{2}$ then we
have maximally efficient dense coding. Thus all those physical systems
where dense coding is possible can be used for encoding of information
by one party and thus all such systems can be used for DSQC. It is
a sufficient criterion for DSQC but not essential (as for cases where
$m=n,$ there dense coding will not happen but encoding will happen).
But dense coding is not even sufficient for quantum dialogue protocol
of Ba An type. Here the demand is more because the encoding is done
by both Alice and Bob using same set of operators. So after the encoding
operation of Alice (say $U_{j})$ and that of Bob (say $U_{i}$) the
final states must also be a member of the mutually orthogonal states
to ensure the deterministic discrimination of the state and thus to
decode the encoded message. To be precise, $|\psi\rangle_{final}=U_{j}U_{i}|\phi_{0}\rangle=U_{j}|\phi_{i}\rangle\in\{|\phi_{0}\rangle,|\phi_{1}\rangle,....,|\phi_{i}\rangle,...,|\phi_{2^{n}-1}\rangle\forall i,j\in\{0,1,...,2^{n}-1\}\Rightarrow U_{B}U_{A}\in\{U_{0},U_{1},U_{2},...,U_{2^{n}-1}\}$.
Thus the product of any two arbitrary unitary operators should be
a member of the set of unitary operators. This is a property of group.
So we may conclude that if we have a set of mutually orthogonal n-qubit
states $\{|\phi_{0}\rangle,|\phi_{1}\rangle,....,|\phi_{i}\rangle,...,|\phi_{2^{n}-1}\rangle\}$
and a set of $m-qubit$ unitary operators $\{U_{0},U_{1},U_{2},...,U_{2^{n}-1}\}$
such that the $U_{i}|\phi_{0}\rangle=|\phi_{i}\rangle$ and $\{U_{0},U_{1},U_{2},...,U_{2^{n}-1}\}$
forms a group under multiplication then it would be sufficient to
construct a quantum dialogue protocol of Ba An type. This is true
in general as information splitting in the sense of Ba An protocol
is possible and Eve knows only $|\psi\rangle_{initial}$ and $|\psi\rangle_{final}$.
Consequently, Eve knows the product of the operators of Alice $U_{j}$
and that of Bob $U_{i}$. Say $U_{j}U_{i}=U_{k}$ and $|\psi\rangle_{final}=U_{k}|\psi\rangle_{initial}$.
Now from the rearrangement theorem of groups we know that each row
and each column in the group multiplication table lists each of the
group elements once and only once. From this, it follows that $U_{k}$
can be decomposed in $2^{n}$ different ways. Thus all possible $2^{n}$
encoding of Bob may lead to the same $U_{k}.$  Now if Eve wants to  obtain the secret information encoded by Alice and Bob then she has to guess either $U_{i}$ or $U_{j}$, i.e. she has to guess among $2^{n}$ equiprobable events. Clearly, the probability of her success is $2^{-n}$ and consequently the quantum dialogue protocol of the present type are more secured when a multi-partite state is used. To be precise, if the quantum dialogue is implemented using Bell state, 5-qubit Brown state and 6-qubit cluster states respectively then the probability of Eve's success is $25\%$, $3.1\%$ and $1.6\%$ respectively. Now since the multi-partite entangled states can be experimentally prepared (for example 6-qubit  cluster state is experimentally prepared by C. Y. Lu \emph{et al.} \cite{6qubit-cluster}) so efficient quantum dialogue protocols with negligible information leakage can be  designed.
Actually, to correctly decompose
$U_{k}$ and thus to decode the encoded information one needs one
of the factor $U_{i}$ or $U_{j}$ and those are available with Alice
and Bob respectively. Consequently, they can successfully decompose
$U_{k}$ and decode each others secret message. Thus the intrinsic
beauty of these protocol lies in the group structure of the unitary
operators and the security (more precisely an upper bound on the security) of the protocol arises in general from
the clever use of rearrangement theorem.

We have obtained a
sufficient condition for construction of quantum dialogue protocol
of Ba An type. It can be noted that all those cases where dense coding
is already reported and the operators used for dense coding form a
group can be used for quantum dialogue. In addition new examples of
possible implementations of dense coding and quantum dialogue can be
obtained in a systematic way. We will provide several examples of
such cases in Section \ref{group construction}. We have discussed the information splitting
process involved in the quantum dialogue protocols in general but
have not described the security of the protocol against intercept-resend
attack. Same would be described in next section while we describe
the generalized protocol. Before we do so, we would like to note something
more on the structure of the unitary operators.

As long as we restrict ourselves to the domain of discrete communication
and claim that the output states are mutually orthogonal, our operation
on individual qubits can only be done by doing nothing (i.e. applying
$I$), flipping the qubit (i.e. applying $X=\sigma_{x}$), flipping
the phase (i.e. applying $Z=\sigma_{z}$) or flipping both phase and
bit (i.e. applying $iY=i\sigma_{y}$). We can not have any other operations,
because that would always take us out of the set of mutually orthogonal
states. Thus no operator other than $(I,\,\sigma_{x},\, i\sigma_{y},\,\sigma_{z})$
are applied to a single qubit and consequently we may decompose any
$m$ qubit unitary operations as tensor product of $m$ single qubit
operations as: \begin{equation}
U(m)=U_{1}(1)\otimes U_{2}(1)\otimes\cdots\otimes U_{m}(1):U_{i}(1)\in\{I,\sigma_{x},i\sigma_{y},\sigma_{z}\},\label{eq:product os single qubitop}\end{equation}
 where $U(j)$ denotes a $j$-qubit unitary operation. Now we are
ready to describe our generalized protocol.


\section{Generalized protocol of quantum dialogue}\label{general protocol}

Our generalized protocol works using the above mentioned $n-qubit$
mutually orthogonal states $\{|\phi_{0}\rangle,|\phi_{1}\rangle,....,|\phi_{i}\rangle,...,|\phi_{2^{n}-1}\rangle\}$
and $m-$qubit unitary operators $\{U_{0},U_{1},...,U_{2^{n}-1}\}$
as follows:
\begin{enumerate}
\item Bob prepares a large number of copies (say $N$ copies) of state $|\phi_{0}\rangle$,
and encodes his classical secret message by applying $m$-qubit unitary
operators $\{U_{0},U_{1}...,U_{2^{n}-1}\}.$ For example, to
encode $0_{1}0_{2}\cdots0_{n},0_{1}0_{2}\cdots1_{n},0_{1}0_{2}\cdots1_{n-1}0_{n},\cdots\cdots\cdots,\\ 1_{1}1_{2}\cdots1_{n}$
he applies $U_{0},U_{1},U_{2},...,U_{2^{n}-1}$ respectively. The
information encoded states should be mutually orthogonal to each other
as discussed above.
\item There are two possibilities: i) $m<n$ i.e. dense coding is possible
and ii) $m=n$ i.e. dense coding is not possible for set of quantum
states and set of unitary operators used for encoding. If $m<n$,
then Bob uses the $m$ photons on which encoding is done as travel
photons and remaining $n-m$ photon as home photon and keeps them
with himself in an ordered sequence $P_{B}=[p_{1}(h_{1},h_{2},...,h_{n-m}),p_{2}(h_{1},h_{2},...,h_{n-m}),...,p_{N}(h_{1},\\ h_{2},...,h_{n-m})]$
where the subscript $1,2,...,N$ denotes the order of a $n-$partite
state $p_{i}=\{h_{1},h_{2},...,h_{n-m},t_{1},t_{2},\cdots,t_{m}\},$
which is in one of the $n$-partite state $|\phi_{j}\rangle$ (value
of $j$ depends on the encoding). Symbol $h$ and $t$ are used to
indicate home photon ($h$) and travel photon ($t$)
respectively. If dense coding is not possible then he has to use all
qubits as travel qubits. In general, he uses all the travel photons
to prepare an ordered sequence $P_{A}=[p_{1}(t_{1},t_{2},\cdots,t_{m}),p_{2}(t_{1},t_{2},\cdots,t_{m}),\\ ...,p_{N}(t_{1},t_{2},\cdots,t_{m})]$.
Now before transmitting the travel qubits to Alice, Bob first prepares
$Nm$ decoy photons in a random sequence of $\{|0\rangle,|1\rangle,|+\rangle,|-\rangle\}$,
i.e. the decoy photon state is $\otimes_{j=1}^{m}|P_{j}\rangle,|P_{j}\rangle\epsilon\{|0\rangle,|1\rangle,|+\rangle,|-\rangle\},(j=1,2,....,m).$
Then Bob reorders the sequence $P_{A}$ of the travel qubits (the
actual ordering is known to Bob only) and inserts $Nm$ decoy photons%
\footnote{When $2x$ qubits (a random mix of message qubits and decoy qubits)
travel through a channel accessible to Eve and x of them are test
for eavesdropping then for any $\delta>0,$ the probability of obtaining
less than $\delta n$ errors on the check qubits (decoy qubits), and
more than $(\delta+\epsilon)n$ errors on the remaining $x$ qubits
is asymptotically less than $\exp[-O(\epsilon^{2}x]$ for large value of x
\cite{Nielsenchuang}. As the unconditional security obtained
in quantum cryptographic protocol relies on the fact that any attempt
of Eavesdropping can be identified. Thus to obtain an unconditional
security we always need to check half of travel qubits for eavesdropping.
Thus we have to randomly add decoy qubits whose number would be equal
to the total number of travel qubits. %
}
, randomly in them and makes a new sequence $P_{A}^{\prime}$ which
contains $2Nm$ photons ($Nm$ travel photons and $Nm$ decoy photons)%
\footnote{Rearrangement of particle orders provide additional security but it
may be avoided in some particular dense coding based cases. Still we
prefer to include it in the protocol as it is essential for the security
in general, specially for cases where $n=m$, because in those cases
since the entire state is sent over the channel a measurement in $\{|\phi_{0}\rangle,|\phi_{1}\rangle,....,|\phi_{i}\rangle,...,|\phi_{2^{n}-1}\rangle\}$
basis by Eve can reveal the encoding of Alice without being detected.
Further, Alice and Bob must be able to ensure that the secret message
do not leak to Eve before she is detected. If we only use the decoy
qubit technique then a considerably large amount of information may
be leaked to Eve. For example, if Bob use $U_{0}=I\otimes I,\, U_{1}=I\otimes i\sigma_{y},\, U_{2}=\sigma_{x}\otimes I,\, U_{3}=\sigma_{x}\otimes i\sigma_{y}$
(It is easy to check that these operators forms a group) to encode
information on 4 qubit $W$ state, $\frac{1}{2}\left(|0001\rangle+|0010\rangle+|0100\rangle+|1000\rangle\right)$
then 75\% of the Bob's message will be leaked to Eve before the measurement
on decoy photons will be successful to detect her \cite{the:high-capacity-wstate}.
Rearrangement of order of particle will avoid such incidents.%
}
 and sends the reordered sequence $P_{A}^{\prime}$ to Alice.
\item After confirming that Alice has received all the $2Nm$ photons, Bob
announces the position of the decoy photons. Alice measures the corresponding
particles in the sequence $P_{A}^{\prime}$ by using $X$ basis or
$Z$ basis at random, here $X=\{|+\rangle,|-\rangle\}$ and $Z=\{|0\rangle,|1\rangle\}$.
After measurement, Alice publicly announces the result of her measurement
and the basis used for the measurement. Now the initial state of the
decoy photon as noted by Bob during preparation and the measurement
outcome of Alice should coincide in all such cases where Alice has
used the same basis as was used to prepare the decoy photon. Bob can
compute the error rate and check whether it exceeds the predeclared
threshold or not. If it exceeds the threshold, then Alice and Bob
abort this communication and repeat the procedure from the beginning.
Otherwise they go on to the next step. So all intercept-resend attacks
will be detected in this step and even if eavesdropping has happened
Eve has no information about the encoding operation executed by Bob
as the sequence is rearranged.
\item Bob announces the actual order.
\item After knowing the actual order, Alice transforms the sequence into
actual order and encodes her information using the same encoding scheme
as was used by Bob. That creates a new sequence $P_{A}^{\prime\prime}$.
Alice prepares $Nm$ decoy photons in a random sequence of $\{|0\rangle,|1\rangle,|+\rangle,|-\rangle\}$,
reorders $P_{A}^{\prime\prime}$ and randomly inserts decoy photon
in that to convert that into a new sequence $P_{A}^{\prime\prime\prime}.$
She then send the sequence $P_{A}^{\prime\prime\prime}$ to Bob.
\item After confirming that Bob has received all the $2Nm$ photons, Alice
announces the position of the decoy photons. Bob measures the corresponding
particles in the sequence $P_{A}^{\prime\prime\prime}$ by using $X$
basis or $Z$ basis at random, here $X=\{|+\rangle,|-\rangle\}$ and
$Z=\{|0\rangle,|1\rangle\}$. After measurement, Bob publicly announces
the result of his measurement and the basis used for the measurement.
Now the initial state of the decoy photon as noted by Alice during
preparation and the measurement outcome of Bob should coincide in
all such cases where Bob has used the same basis as was used to prepare
the decoy photon. Alice can compute the error rate and check whether
it exceeds the predeclared threshold or not. If it exceeds the threshold,
then Alice and Bob abort this communication and repeat the procedure
from the beginning. Otherwise they go on to the next step. These makes
the protocol safe from all kind of eavesdropping strategy in the return
path.
\item Alice announces the actual order.
\item Bob reorders the sequence to obtain $P_{A}^{\prime\prime}$. Recombines
it with $P_{B}$ and measures each $n-$partite state in $\{|\phi_{0}\rangle,|\phi_{1}\rangle,....,|\phi_{i}\rangle,...,|\phi_{2^{n}-1}\rangle\}$
basis. As he already knows the unitary operators applied by him or
the state $|\phi_{i}\rangle$ sent by him, he can now easily decode
the message encoded by Alice. After the measurement Bob publicly announces
the final states that he has obtained in sequence.
\item Now as Alice knows her encoding into a particular state she will be
able to decode the secret message of Bob.
\end{enumerate}
Now we would like to note that in case $m=n$ then in Step 5 after
knowing the actual order Alice could have decoded the message of Bob
and in that case the public announcement of Bob in Step 8 and the
entire Step 9 would be redundant. In such case, the protocol essentially
get decomposed into two protocols of DSQC: one from Alice to Bob and
the other from Bob to Alice. That is not really in accordance with
the true spirit of the quantum dialogue protocols and consequently
we have excluded such cases from the remaining discussion. It is straight
to note that neither DSQC nor quantum dialogue protocol requires dense coding
as an essential resource but it is always useful. In case of DSQC
it is sufficient \cite{Anindita} and in case of quantum dialogue,
in addition, the unitary operators should form a group. Now to further
clarify that dense coding is not essential for a quantum dialogue protocol
we may give a different example, where only the subset of the
basis set is used. To be precise, we may use $U_{0}=I\otimes I,\, U_{1}=I\otimes i\sigma_{y},\, U_{2}=\sigma_{x}\otimes I,\, U_{3}=\sigma_{x}\otimes i\sigma_{y}$
as our unitary operators (these operators form a group) and following
4-qubit $W$ states provide example of required orthogonal states
\cite{the:high-capacity-wstate}: \begin{equation}
\begin{array}{lcl}
|\phi_{0}\rangle & = & U_{0}|\phi_{0}\rangle=\frac{1}{2}\left(|0001\rangle+|0010\rangle+|0100\rangle+|1000\rangle\right),\\
|\phi_{1}\rangle & = & U_{1}|\phi_{0}\rangle=\frac{1}{2}\left(|0000\rangle-|0011\rangle-|0101\rangle-|1001\rangle\right),\\
|\phi_{2}\rangle & = & U_{2}|\phi_{0}\rangle=\frac{1}{2}\left(|0011\rangle+|0000\rangle+|0110\rangle+|1010\rangle\right),\\
|\phi_{3}\rangle & = & U_{3}|\phi_{0}\rangle=\frac{1}{2}\left(|0010\rangle-|0001\rangle-|0111\rangle-|1011\rangle\right).\\ \end{array}\label{eq:pustak1}\end{equation}
This resource would be good enough for an efficient protocol of quantum
dialogue. But here we can send only 2 bits of classical information
using 2 ebits so this is not dense coding. Consequently, we may conclude
that dense coding is not essential for quantum dialogue. Thus the fact
that the unitary operators forms a group is most crucial property.
In addition since we are interested in discrete variable quantum communication
our $m$ qubit unitary operators are supposed to be formed as product
of single qubit Pauli operators (\ref{eq:product os single qubitop}).
Let us now try to construct such groups of unitary operators that
will be useful for quantum dialogue.


\section{How to construct groups and subgroups of unitary operators required
for quantum dialogue?}\label{group construction}

We have seen that the set of unitary operations used for encoding
of information are required to form a group under multiplication (without
global phase). Since the Hilbert space of $n-$qubit system is $C^{2^{n}}$
so we have $2^{n}$ mutually orthogonal state vectors%
\footnote{We may use a subset of these state vectors for encoding but the measurement
is done using the entire basis set and any such use of subspace always
reduce efficiency.%
} and $2^{n}$ unitary operators. The exclusion of continuous variable
communication ensures that the unitary operators are formed by combining
Pauli operators (more precisely by combining $I,\,\sigma_{x},\, i\sigma_{y},\,\sigma_{z})$.
Now it is straight forward to observe that $\{I,\,\sigma_{x},\, i\sigma_{y},\,\sigma_{z}\}$
forms a group under multiplication. This is so because there exist
an identity operator $I.$ Further the product of two Pauli matrices
are described as \begin{equation}
\sigma_{i}.\sigma_{j}=I\delta_{i,j}+i\sum_{k}\epsilon_{ijk}\sigma_{k},\label{eq:pauli-product}\end{equation}
where $\epsilon_{ijk}$ is the Levi-Civita symbol. Thus all Pauli
operators are self inverse and consequently inverse of each element
of the group is also a member of the group. Since these unitary operators
are matrices they automatically satisfy associativity and finally
since the product of any two of these operators is another element
of the set (see the group multiplication table as provided in Table
\ref{table_groupmultiplication_g1}) so it forms a group of order
4. Here we would like to note that the multiplication rule is defined
as bit wise dot product in such a way that the global phase (a common
minus sign in the product matrix) is ignored which is consistent with
the quantum mechanics. Thus $G_{1}=\{I,\,\sigma_{x},\, i\sigma_{y},\,\sigma_{z}\}=\{I,\, X,\, iY,\, Z\}$
forms a group of order 4 under multiplication where the elements are
one-qubit unitary operators. Further under the above defined multiplication
rule the group is Abelian. %
\begin{table}
\caption{\label{table_groupmultiplication_g1}Group multiplication table for $G_{1}=\{I,\,\sigma_{x},\, i\sigma_{y},\,\sigma_{z}\}=\{I,\, X,\, iY,\, Z\}$. }
\begin{ruledtabular}
\begin{tabular}{ccccc}

Pauli Operators & $I$ & $\sigma_{x}$ & $i\sigma_{y}$ & $\sigma_{z}$\tabularnewline
\hline
$I$ & $I$ & $\sigma_{x}$ & $i\sigma_{y}$ & $\sigma_{z}$\tabularnewline

$\sigma_{x}$ & $\sigma_{x}$ & $I$ & $\sigma_{z}$ & $i\sigma_{y}$\tabularnewline

$i\sigma_{y}$ & $i\sigma_{y}$ & $\sigma_{z}$ & $I$ & $\sigma_{x}$\tabularnewline

$\sigma_{z}$ & $\sigma_{z}$ & $i\sigma_{y}$ & $\sigma_{x}$ & $I$\tabularnewline

\end{tabular}
\end{ruledtabular}
\end{table}

 Now since $G_{1}$ is a group of order 4 so \begin{equation}
\begin{array}{ccc}
G_{2} & = & G_{1}\otimes G_{1}=\{I,\, X,\, iY,\, Z\}\otimes\{I,\, X,\, iY,\, Z\}\\
 & = &  I\otimes I,\, I\otimes X,\, I\otimes iY,\, I\otimes Z,\, X\otimes I,\,X\otimes X, \\
  &  &  X\otimes iY,\, X\otimes Z,\,iY\otimes I,\, iY\otimes X,\, iY\otimes iY,\\
 &  &  iY\otimes Z,Z\otimes I,\, Z\otimes X,\, Z\otimes iY,\, Z\otimes Z. \end{array}\label{eq:pustak2}\end{equation}
is a group of order $16=2^{4}.$ In general, if we have a set of all
$n$-qubit unitary operations such that each $n$-qubit operation
can be viewed as tensor product of $n$ single qubit operations, which
are elements of $G_{1}$, then these unitary operators form a group
$G_{n}$ of order $2^{2^{n}}=4^{n}$ or in other words $G_{n}=G_{1}^{\otimes n}$
is a group. Now we may recall the well known first Sylow theorem,
which is stated as follows:

\textbf{Sylow Theorem:} If $G$ is a group of order $p^{k}m$, with
$p$ prime, and $(p,m)=1$ , then $G$ contains a subgroup of order
$p^{i}$ for each $i\leq k$ and every subgroup of $G$ of order $p^{i}$,
is normal in some subgroup of order $p^{i+1}$. In particular, Sylow
$p-$subgroups exist.

So in our case, $p=2,\, m=1,\, k=2^{n}$ and in general $G_{n}$ has
Sylow 2-subgroups of order $2,\,4,\,8,...2^{2^{n}}$. Thus $G_{2}$
has subgroups of order $16,\,8,\,4,\,2,\,1.$ An important question
here is how many such subgroups are there? Unfortunately the third
Sylow theorem provides an answer to this question for the largest
subgroup only. In our case the largest subgroup is the group $G_{n}$
itself and consequently it has only one subgroup of order $2^{2^{n}}.$
But we are interested about subgroups of order $2^{i}$ where $0<i<2^{n}$
in general and specially about subgroups of order $2^{n-1}.$ The
reason behind this specific interest would be clear if you consider
the odd qubit cases. To be precise, let us think that we have set
of $2^{t}$ mutually orthogonal $t$-qubit states where $t$ is odd.
Now for efficient encoding in this system one needs $2^{t}$ unitary
operators. So these operators will be at least $\frac{t+1}{2}$ qubit
operators and thus a subset of group $G_{n}:t\geq n\geq\frac{t+1}{2}.$
For example, let us consider the case of 3 qubit $GHZ$ state. Here
we have $2^{3}=8$ basis vectors and for successful encoding we will require
an order 8 group of unitary operators. Each operator must be at least
$2$-qubit operator because if you use only single qubit operation
then at most you can encode 4 alternatives (2-bit information). Thus
the group to be used here must be an order 8 subgroup of $G_{2}$ or
some higher order group (say $G_{3}, G_{4},$ ...etc.) as they can
also have order 8 subgroups. For example, all subgroups of $G_{2}$
will be subgroups of $G_{3}$ with third element being $I.$ But since
we have a three qubit system we can not use 4 or more qubit operations
on them and these limits the possible subgroups that can be used for
encoding operation suitable for quantum dialogue. Thus it is clear
that the order 8 subgroups of $G_{2}$ will play important role in
the designing of quantum dialogue protocol using 3-qubit states (such
as $GHZ$ state, $GHZ$-\emph{like} state etc.) and in some four qubit
states where complete dense coding is not possible (such as 4-qubit
$W$ state, 4 qubit Cluster state etc.). This is why we are specially
interested about the order 8 subgroups of $G_{2}$. We can construct
few subgroups very easily. For example, since each Pauli gates are
self inverse so $\{I,\, X\},\{I,\, iY\},\{I,\, Z\}$ are subgroups
of $G_{1}$ consequently, following are order 8 subgroups of $G_{2}$:
\begin{equation}
\begin{array}{ccc}
G_{2}^{1}(8) & = & \{I,\, X,\, iY,\, Z\}\otimes\{I,\, X\} \\
 &= & \{I\otimes I,\, X\otimes I,\, iY\otimes I,\, Z\otimes I,\,\\
  & & I\otimes X,\, X\otimes X,\, iY\otimes X,\, Z\otimes X\}\\
G_{2}^{2}(8) & = & \{I,\, X,\, iY,\, Z\}\otimes\{I,\, iY\} \\
&= & \{I\otimes I,\, X\otimes I,\, iY\otimes I,\, Z\otimes I,\,\\
  & &I\otimes iY,\, X\otimes iY,\, iY\otimes iY,\, Z\otimes iY\}\\
G_{2}^{3}(8) & = & \{I,\, X,\, iY,\, Z\}\otimes\{I,\, Z\} \\
& = & \{I\otimes I,\, X\otimes I,\, iY\otimes I,\, Z\otimes I,\,\\
  & &I\otimes Z,\, X\otimes Z,\, iY\otimes Z,\, Z\otimes Z\}\\
G_{2}^{4}(8) & = & \{I,\, X\}\otimes\{I,\, X,\, iY,\, Z\} \\
& = & \{I\otimes I,\, I\otimes X,\, I\otimes iY,\, I\otimes Z,\,\\
  & &X\otimes I,\, X\otimes X,\, X\otimes iY,\, X\otimes Z\}\\
G_{2}^{5}(8) & = & \{I,\, iY\}\otimes\{I,\, X,\, iY,\, Z\} \\
& = & \{I\otimes I,\, I\otimes X,\, I\otimes iY,\, I\otimes Z,\,\\
  & & iY\otimes I,\, iY\otimes X,\, iY\otimes iY,\, iY\otimes Z\}\\
G_{2}^{6}(8) & = & \{I,Z\}\otimes\{I,\, X,\, iY,\, Z\} \\
& = & \{I\otimes I,\, I\otimes X,\, I\otimes iY,\, I\otimes Z,\,\\
  & & Z\otimes I,\, Z\otimes X,\, Z\otimes iY,\, Z\otimes Z\}\end{array}\label{eq:pustak3}\end{equation}
where $G_{n}^{j}(m)$ denotes $j^{th}$ subgroup of order $m<4^{n}$
of the group $G_{n}$ whose order is $4^{n}.$ Similarly we may construct
$3n$ subgroups of order $2^{2^{n}-1}$ of $G_{n}$ as follows: \begin{equation}\begin{array}{c}
G_{1}^{\otimes i}\otimes\left\{ I,X\right\} \otimes G_{1}^{n-i-1},\\G_{1}^{\otimes i}\otimes\left\{ I,iY\right\} \otimes G_{1}^{n-i-1},\\
G_{1}^{\otimes i}\otimes\left\{ I,Z\right\} \otimes G_{1}^{n-i-1}\label{eq:pustak4}\end{array}\end{equation}
where $i$ varies from $0$ to $n-1.$ For example, we can easily
obtain following $9$ order $32$ subgroups of $G_{3}$:
\begin{equation}
\begin{array}{c}
G_{3}^{1}(32)=G_{2}\otimes\{I,X\},\\
G_{3}^{2}(32)=G_{2}\otimes\{I,iY\},\\
G_{3}^{3}(32)=G_{2}\otimes\{I,Z\},\\
G_{3}^{4}(32)=\{I,X\}\otimes G_{2},\\
G_{3}^{5}(32)=\{I,iY\}\otimes G_{2},\\
G_{3}^{6}(32)=\{I,Z\}\otimes G_{2},\\
G_{3}^{7}(32)=G_{1}\otimes\left\{ I,X\right\} \otimes G_{1},\\
G_{3}^{8}(32)=G_{1}\otimes\left\{ I,iY\right\} \otimes G_{1},\\
G_{3}^{9}(32)=G_{1}\otimes\left\{ I,Z\right\} \otimes G_{1}.
\end{array}\label{eq:subgroups of g3}\end{equation}
 But the above set in (\ref{eq:pustak3}) is not complete. For example  following
order 8 subgroups of $G_{2}$,  are not in that set,

\begin{equation}
\begin{array}{ccc}
G_{2}^{7}(8)&=&{ I\otimes I, I\otimes Z, Z\otimes I, Z\otimes Z,}
\\
& & {X\otimes X, iY\otimes X, X\otimes iY, iY\otimes iY}
\\
G_{2}^{8}(8) & = & { I\otimes I, Z\otimes Z, X\otimes iY, iY\otimes X,}
\\
& & {I\otimes X,\, Z\otimes iY,\, iY\otimes I,\, X\otimes Z}
\\
G_{2}^{9}(8) & = & { I\otimes I, Z\otimes Z, X\otimes iY, iY\otimes X,}
\\
& & { X\otimes I, iY\otimes Z, Z\otimes X, I\otimes iY}
\\
G_{2}^{10}(8) & = & {I\otimes I, X\otimes I, I\otimes X, X\otimes X}
\\
& & {Z\otimes Z, iY\otimes Z, Z\otimes iY, iY\otimes iY}
\\
G_{2}^{11}(8) & = & {I\otimes I, iY\otimes I, I\otimes iY, iY\otimes iY,}
\\
& & {Z\otimes Z, Z\otimes X, X\otimes Z,X\otimes X}
\\ \end{array}.\label{eq:pustak5}\end{equation}

Now if we try to implement a quantum dialogue protocol with 2 qubit
EPR states as was done in the original protocol of Ba An then the
single qubit unitary operations defined in $G_{1}$ would be sufficient.
Similarly as the elements of $G_{2}$ does dense coding for 4 qubit
Cluster states and 4 qubit $\Omega$ states (as shown in Table \ref{table:omega_cluster})
it would be straight forward to note that we can have quantum dialogue
protocol for 4-qubit Cluster states and 4 qubit $\Omega$ state.
%
\begin{table*}
\caption{\label{table:omega_cluster}Dense coding operation for 4 qubit Omega states $\Omega$ and Cluster
states $|C\rangle$. The unitary operators operate on the qubits 1
and 3 in both the cases. As all the elements of $G_{2}$ group are
used as unitary operators for encoding, these two states may be used
for quantum dialogue using the generalized protocol presented here.}
\begin{ruledtabular}\begin{tabular}{ccc}

Unitary Operations & $|\Omega\rangle_{0}=\frac{1}{2}(|0000\rangle+|0110\rangle+|1001\rangle-|1111\rangle)$ & $|C\rangle_{0}=\frac{1}{2}(|0000\rangle+|0011\rangle+|1100\rangle-|1111\rangle)$\tabularnewline
\hline
$U_{0}=I\otimes I$ & $\frac{1}{2}(|0000\rangle+|0110\rangle+|1001\rangle-|1111\rangle)$ & $\frac{1}{2}(|0000\rangle+|0011\rangle+|1100\rangle-|1111\rangle)$\tabularnewline

$U_{1}=I\otimes Z$ & $\frac{1}{2}(|0000\rangle-|0110\rangle+|1001\rangle+|1111\rangle)$ & $\frac{1}{2}(|0000\rangle-|0011\rangle+|1100\rangle+|1111\rangle)$\tabularnewline

$U_{2}=Z\otimes I$ & $\frac{1}{2}(|0000\rangle+|0110\rangle-|1001\rangle+|1111\rangle)$ & $\frac{1}{2}(|0000\rangle+|0011\rangle-|1100\rangle+|1111\rangle)$\tabularnewline

$U_{3}=Z\otimes Z$ & $\frac{1}{2}(|0000\rangle-|0110\rangle-|1001\rangle-|1111\rangle)$ & $\frac{1}{2}(|0000\rangle-|0011\rangle-|1100\rangle-|1111\rangle)$\tabularnewline

$U_{4}=I\otimes X$ & $\frac{1}{2}(|0010\rangle+|0100\rangle+|1011\rangle-|1101\rangle)$ & $\frac{1}{2}(|0001\rangle+|0010\rangle-|1101\rangle+|1110\rangle)$\tabularnewline

$U_{5}=I\otimes iY$ & $\frac{1}{2}(-|0010\rangle+|0100\rangle-|1011\rangle-|1101\rangle)$ & $\frac{1}{2}(-|0001\rangle+|0010\rangle+|1101\rangle+|1110\rangle)$\tabularnewline

$U_{6}=Z\otimes X$ & $\frac{1}{2}(|0010\rangle+|0100\rangle-|1011\rangle+|1101\rangle)$ & $\frac{1}{2}(|0001\rangle+|0010\rangle+|1101\rangle-|1110\rangle)$\tabularnewline

$U_{7}=Z\otimes iY$ & $\frac{1}{2}(-|0010\rangle+|0100\rangle+|1011\rangle+|1101\rangle)$ & $\frac{1}{2}(-|0001\rangle+|0010\rangle-|1101\rangle-|1110\rangle)$\tabularnewline

$U_{8}=X\otimes I$ & $\frac{1}{2}(|1000\rangle+|1110\rangle+|0001\rangle-|0111\rangle)$ & $\frac{1}{2}(|0100\rangle-|0111\rangle+|1000\rangle+|1011\rangle)$\tabularnewline

$U_{9}=X\otimes Z$ & $\frac{1}{2}(|1000\rangle-|1110\rangle+|0001\rangle+|0111\rangle)$ & $\frac{1}{2}(|0100\rangle+|0111\rangle+|1000\rangle-|1011\rangle)$\tabularnewline

$U_{10}=iY\otimes I$ & $\frac{1}{2}(-|1000\rangle-|1110\rangle+|0001\rangle-|0111\rangle)$ & $\frac{1}{2}(|0100\rangle-|0111\rangle-|1000\rangle-|1011\rangle)$\tabularnewline

$U_{11}=iY\otimes Z$ & $\frac{1}{2}(-|1000\rangle+|1110\rangle+|0001\rangle+|0111\rangle)$ & $\frac{1}{2}(|0100\rangle+|0111\rangle-|1000\rangle+|1011\rangle)$\tabularnewline

$U_{12}=X\otimes X$ & $\frac{1}{2}(|1010\rangle+|1100\rangle+|0011\rangle-|0101\rangle)$ & $\frac{1}{2}(-|0101\rangle+|0110\rangle+|1001\rangle+|1010\rangle)$\tabularnewline

$U_{13}=X\otimes iY$ & $\frac{1}{2}(-|1010\rangle+|1100\rangle-|0011\rangle-|0101\rangle)$ & $\frac{1}{2}(|0101\rangle+|0110\rangle-|1001\rangle+|1010\rangle)$\tabularnewline

$U_{14}=iY\otimes X$ & $\frac{1}{2}-|1010\rangle-|1100\rangle+|0011\rangle-|0101\rangle)$ & $\frac{1}{2}(-|0101\rangle+|0110\rangle-|1001\rangle-|1010\rangle)$\tabularnewline

$U_{15}=iY\otimes iY$ & $\frac{1}{2}(|1010\rangle-|1100\rangle-|0011\rangle-|0101\rangle)$ & $\frac{1}{2}(|0101\rangle+|0110\rangle+|1001\rangle-|1010\rangle)$\tabularnewline

\end{tabular}
\end{ruledtabular}
\end{table*}

\begin{table}
\caption{\label{table_GHZ_g_1_2_8}Dense coding of $GHZ$ states using the elements of $G_{2}^{1}(8).$}
\begin{ruledtabular}
\begin{tabular}{cc}

Unitary operations  & $GHZ$ states\tabularnewline
on qubits 1 and 2 & \tabularnewline \hline
$U_{0}=I\otimes I$ & $\frac{1}{\sqrt{2}}(|000\rangle+|111\rangle)$\tabularnewline

$U_{1}=Z\otimes I$ & $\frac{1}{\sqrt{2}}(|000\rangle-|111\rangle)$\tabularnewline

$U_{2}=X\otimes I$ & $\frac{1}{\sqrt{2}}(|100\rangle+|011\rangle)$\tabularnewline

$U_{3}=iY\otimes I$ & $\frac{1}{\sqrt{2}}(-|100\rangle+|011\rangle)$\tabularnewline

$U_{4}=I\otimes X$ & $\frac{1}{\sqrt{2}}(|010\rangle+|101\rangle)$\tabularnewline

$U_{5}=Z\otimes X$ & $\frac{1}{\sqrt{2}}(|010\rangle-|101\rangle)$\tabularnewline

$U_{6}=X\otimes X$ & $\frac{1}{\sqrt{2}}(|110\rangle+|001\rangle)$\tabularnewline

$U_{7}=iY\otimes X$ & $\frac{1}{\sqrt{2}}(-|110\rangle+|001\rangle)$\tabularnewline

\end{tabular}
\end{ruledtabular}
\end{table}

\begin{table}
\caption{\label{table:GHZ_2_2_8}Dense coding of GHZ states using the elements of $G_{2}^{2}(8).$}
\begin{ruledtabular}
\begin{tabular}{cc}

Unitary  operators on  & $GHZ$ states\tabularnewline
 qubits 1 and 2 &  \tabularnewline \hline
$U_{0}$ = $I\otimes I$ & $\frac{1}{\sqrt{2}}(|000\rangle+|111\rangle)$\tabularnewline

$U_{1}$ = $Z\otimes I$ & $\frac{1}{\sqrt{2}}(|000\rangle-|111\rangle)$\tabularnewline

$U_{2}$ = $X\otimes I$ & $\frac{1}{\sqrt{2}}(|100\rangle+|011\rangle)$\tabularnewline

$U_{3}$ = $iY\otimes I$ & $\frac{1}{\sqrt{2}}(-|100\rangle+|011\rangle)$\tabularnewline

$U_{4}$ = $I\otimes iY$ & $\frac{1}{\sqrt{2}}(-|010\rangle+|101\rangle)$\tabularnewline

$U_{5}$ = $Z\otimes iY$ & $\frac{1}{\sqrt{2}}(-|010\rangle-|101\rangle)$\tabularnewline

$U_{6}$ = $X\otimes iY$ & $\frac{1}{\sqrt{2}}(-|110\rangle+|001\rangle)$\tabularnewline

$U_{7}$ = $iY\otimes iY$ & $\frac{1}{\sqrt{2}}(|110\rangle+|001\rangle)$\tabularnewline

\end{tabular}
\end{ruledtabular}
\end{table}

\begin{table}
\caption{\label{table:grpmultiplication_g_1_2_8}Group multiplication table for $G_{2}^{1}(8)$, where $U_{0}=I\otimes I$,
$U_{1}=Z\otimes I,$ $U_{2}=X\otimes I$, $U_{3}=iY\otimes I$, $U_{4}=I\otimes X,$
$U_{5}=Z\otimes X$ , $U_{6}=X\otimes X$, $U_{7}=iY\otimes X$. }
\begin{ruledtabular}
\begin{tabular}{ccccccccc}

Unitary operations & $U_{0}$ & $U_{1}$ & $U_{2}$ & $U_{3}$ & $U_{4}$ & $U_{5}$ & $U_{6}$ & $U_{7}$\tabularnewline
\hline
$U_{0}$ & $U_{0}$ & $U_{1}$ & $U_{2}$ & $U_{3}$ & $U_{4}$ & $U_{5}$ & $U_{6}$ & $U_{7}$\tabularnewline

$U_{1}$ & $U_{1}$ & $U_{0}$ & $U_{3}$ & $U_{2}$ & $U_{5}$ & $U_{4}$ & $U_{7}$ & $U_{6}$\tabularnewline

$U_{2}$ & $U_{2}$ & $U_{3}$ & $U_{0}$ & $U_{1}$ & $U_{6}$ & $U_{7}$ & $U_{4}$ & $U_{5}$\tabularnewline

$U_{3}$ & $U_{3}$ & $U_{2}$ & $U_{1}$ & $U_{0}$ & $U_{7}$ & $U_{6}$ & $U_{5}$ & $U_{4}$\tabularnewline

$U_{4}$ & $U_{4}$ & $U_{5}$ & $U_{6}$ & $U_{7}$ & $U_{0}$ & $U_{1}$ & $U_{2}$ & $U_{3}$\tabularnewline

$U_{5}$ & $U_{5}$ & $U_{4}$ & $U_{7}$ & $U_{6}$ & $U_{1}$ & $U_{0}$ & $U_{3}$ & $U_{2}$\tabularnewline

$U_{6}$ & $U_{6}$ & $U_{7}$ & $U_{4}$ & $U_{5}$ & $U_{2}$ & $U_{3}$ & $U_{0}$ & $U_{1}$\tabularnewline

$U_{7}$ & $U_{7}$ & $U_{6}$ & $U_{5}$ & $U_{4}$ & $U_{3}$ & $U_{2}$ & $U_{1}$ & $U_{0}$\tabularnewline

\end{tabular}
\end{ruledtabular}
\end{table}

\begin{table}
\caption{\label{table:groupmutiplication_butnot_densecodng}In $G_{2}^{3}(8)$ dense coding is not possible but group multiplication table exists.
{}``Dense coding is helpful (sufficient) but not essential''.}
\begin{ruledtabular}
\begin{tabular}{cc}

Unitary   operations on  & Non orthogonal \tabularnewline
 qubits 1 and 2 & GHZ states \tabularnewline
\hline
$U_{0}$ = $I\otimes I$ & $\frac{1}{\sqrt{2}}(|000\rangle+|111\rangle)$\tabularnewline

$U_{1}$ = $Z\otimes I$ & $\frac{1}{\sqrt{2}}(|000\rangle-|111\rangle)$\tabularnewline

$U_{2}$ = $X\otimes I$ & $\frac{1}{\sqrt{2}}(|100\rangle+|011\rangle)$\tabularnewline

$U_{3}$ = $iY\otimes I$ & $\frac{1}{\sqrt{2}}(-|100\rangle+|011\rangle)$\tabularnewline

$U_{4}$ = $I\otimes Z$ & $\frac{1}{\sqrt{2}}(|000\rangle-|111\rangle)$\tabularnewline

$U_{5}$ = $Z\otimes Z$ & $\frac{1}{\sqrt{2}}(|000\rangle+|111\rangle)$\tabularnewline

$U_{6}$ = $X\otimes Z$ & $\frac{1}{\sqrt{2}}(|100\rangle-|011\rangle)$\tabularnewline

$U_{7}$ = $iY\otimes Z$ & $\frac{1}{\sqrt{2}}(-|100\rangle-|011\rangle)$\tabularnewline
\end{tabular}
\end{ruledtabular}
\end{table}

\begin{table}
\caption{\label{table_densecoding_butnot_groupmultiplicatin}Dense coding of $GHZ$-\emph{like} states. Unitary operators do not
form a group.}
\begin{ruledtabular}
\begin{tabular}{cc}

Unitary operator  & State\tabularnewline
\hline
$U_{0}=I\otimes I$  & $\frac{\left|\psi^{+}0\right\rangle +\left|\psi^{-}1\right\rangle }{\sqrt{2}}$\tabularnewline

$U_{1}=X\otimes X$  & $\frac{\left|\psi^{+}0\right\rangle -\left|\psi^{-}1\right\rangle }{\sqrt{2}}$\tabularnewline

$U_{2}=Z\otimes I$  & $\frac{\left|\psi^{-}0\right\rangle +\left|\psi^{+}1\right\rangle }{\sqrt{2}}$\tabularnewline

$U_{3}=iY\otimes I$  & $\frac{\left|\phi^{+}1\right\rangle -\left|\phi^{-}0\right\rangle }{\sqrt{2}}$\tabularnewline

$U_{4}=I\otimes X$  & $\frac{\left|\phi^{+}0\right\rangle +\left|\phi^{-}1\right\rangle }{\sqrt{2}},$\tabularnewline

$U_{5}=X\otimes I$  & $\frac{\left|\phi^{+}0\right\rangle -\left|\phi^{-}1\right\rangle }{\sqrt{2}}$\tabularnewline

$U_{6}=I\otimes iY$  & $\frac{\left|\phi^{-}0\right\rangle +\left|\phi^{+}1\right\rangle }{\sqrt{2}}$\tabularnewline

$U_{7}=iY\otimes X$  & $\frac{\left|\psi^{+}1\right\rangle -\left|\psi^{-}0\right\rangle }{\sqrt{2}}$\tabularnewline

\end{tabular}
\end{ruledtabular}
\end{table}


Now it would be interesting to note that the elements of $G_{2}^{1}(8)$
and $G_{2}^{2}(8)$ may be used for dense coding of $GHZ$ states as
shown in Table \ref{table_GHZ_g_1_2_8} and Table \ref{table:GHZ_2_2_8}
respectively. We have already logically established that $G_{2}^{1}(8)$
is an order 8 subgroup of $G_{2}$ but for the convenience of the
readers we have also provided group multiplication table for $G_{2}^{1}(8)$
as Table \ref{table:grpmultiplication_g_1_2_8}. Similar tables can
easily be constructed for other subgroups and groups mentioned here.
Since this verification is an easy task we have not provided such
tables here. Now since $G_{2}^{1}(8)$ and $G_{2}^{2}(8)$ are groups
of appropriate order and can be used for dense coding in $GHZ$ states
we may easily conclude that $GHZ$ states may be used for quantum
dialogue. Now the permutation symmetry of the $GHZ$ states and these
order $8$ subgroups of $G_{2}$ ensures that $G_{2}^{4}(8)$ and
$G_{2}^{5}(8)$ will also perform dense coding in $GHZ$ states and
consequently be useful for quantum dialogue. Earlier Xia \emph{et
al.} \cite{xia} had provided protocol for quantum dialogue using
$GHZ$ state. The set of unitary operators used by them coincides
with $G_{2}^{1}(8)$. Since their protocol is an example of Ba An
type of protocol, naturally we obtain that as a special case of our
more generalized protocol. In addition we obtain at least three more
different ways to do encoding operations on $GHZ$ states for quantum
dialogue. From the observation that $G_{2}^{1}(8),\, G_{2}^{2}(8),\, G_{2}^{4}(8)$
and $G_{2}^{5}(8)$ are useful for quantum dialogue using $GHZ$ states,
one may be tempted to use $G_{2}^{3}(8)$ or $G_{2}^{6}(8)$ for the
same purpose. But that would not work. Just as an example, in Table
\ref{table:groupmutiplication_butnot_densecodng} we have shown that
$G_{2}^{3}(8)$ can not be used for dense coding operation on $GHZ$
states because $U_{0}\frac{|000\rangle+|111\rangle}{\sqrt{2}}=U_{5}\frac{|000\rangle+|111\rangle}{\sqrt{2}},U_{1}\frac{|000\rangle-|111\rangle}{\sqrt{2}}=U_{4}\frac{|000\rangle-|111\rangle}{\sqrt{2}},U_{2}\frac{|100\rangle+|011\rangle}{\sqrt{2}}=U_{7}\frac{|100\rangle+|011\rangle}{\sqrt{2}}$
and $U_{3}\frac{|100\rangle-|011\rangle}{\sqrt{2}}=U_{6}\frac{|100\rangle-|011\rangle}{\sqrt{2}}.$
Thus the encoded states are not unique (are not linearly independent).
In Table \ref{table:groupmutiplication_butnot_densecodng} we have
described the case where the encoding operations are done on first
two qubits but the permutation symmetry of the $GHZ$ states ensure
that the same conclusion will remain valid even if we apply the same
unitary operations on qubits 2, 3 or 1, 3 and also for the set of
unitary operations $G_{2}^{6}(8)$. Thus this example shows that formation
of a group of unitary operators alone is not sufficient for quantum
dialogue we also need appropriate quantum states on which that particular
group of unitary operators can be applied to implement quantum dialogue.
With the similar intention we may note that there exists a set of
8 operators (as shown in Table \ref{table_densecoding_butnot_groupmultiplicatin})
which may be used for dense coding in $GHZ$-\emph{like} states \cite{Anindita}
but these operators do not form a group under multiplication since
$U_{7}U_{6}=iY\otimes Z,\, U_{6}U_{5}=X\otimes iY$ etc. are not in
the set $\{U_{0},U_{1},\cdots,U_{7}\}$ used here for dense coding.
Consequently, if Bob applies $U_{6}$ and Alice applies $U_{7}$ then
the quantum dialogue protocol will fail. Thus this example shows that
dense coding alone is not also sufficient for quantum dialogue. This
does not really mean that we can not obtain a quantum dialogue protocol
with $GHZ$-\emph{like} state; another set of operators may satisfy
the sufficiency condition introduced here. To establish this point,
we may look at Table \ref{table:densecoding_GHZlikeandW_g_9_2_8}
where it is shown that $G_{2}^{9}(8)$ may be used for dense coding
in $GHZ$-\emph{like} state and also for dense coding in 4 qubit $W$
state. Thus both 3 qubit $GHZ$-\emph{like} state and 4 qubit $W$
state can be used for quantum dialogue. These particular examples
just show that for quantum dialogue we need to simultaneously satisfy
both the conditions discussed above. In Table \ref{table_g_8_2_8}
we have shown an alternative way for encoding of information in 4
qubit $W$ states for implementation of quantum dialogue protocol.
To be precise, Table \ref{table_g_8_2_8} shows that we can also use
the elements of $G_{2}^{8}(8)$ to implement quantum dialogue using
4 qubit $W$ states. Thus there are at least two different ways to
implement quantum dialogue using $W$ states. Similarly there are
at least six different ways to implement quantum dialogue using $GHZ$-\emph{like}
states by using $G_{2}^{2}(8),G_{2}^{3}(8),G_{2}^{5}(8),G_{2}^{6}(8),G_{2}^{8}(8)$
and $G_{2}^{9}(8)$. The dense coding of $GHZ$-\emph{like} states
using the elements of $G_{2}^{9}(8),G_{2}^{3}(8)$ and $G_{2}^{6}(8)$
subgroups are shown in Table \ref{table:densecoding_GHZlikeandW_g_9_2_8},
Table \ref{table_g_3_2_8} and Table \ref{table_g_6_2_8} respectively.
The other three cases (i.e. dense coding using $G_{2}^{2}(8),G_{2}^{5}(8),G_{2}^{8}(8)$)
can be verified in the similar way. Four qubit $Q_{4}$ state and 4
qubit $Q_{5}$ state introduced in \cite{Pradhan-Agarwal-Pati} are
also very interesting as there are at least 2 different ways to implement
quantum dialogue using $Q_{4}$ state and at least 3 different ways
to implement quantum dialogue using $Q_{5}$ state. To be precise,
in Table \ref{table_g_6_2_8} and Table \ref{table_g_7_2_8} we have
shown that quantum dialogue can be implemented using $Q_{4}$ state
by using the elements $G_{2}^{6}(8)$ and $G_{2}^{7}(8)$ respectively.
Similarly,  we can perform dense coding on $Q_{5}$ state  by elements of $G_{2}^{3}(8),\, G_{2}^{4}(8)$
and $G_{2}^{5}(8).$ In Table \ref{table_g_4_2_8} dense coding of
$Q_{5}$ state by using the elements of $G_{2}^{4}(8)$ is shown.
The other two examples can be verified in the same manner. Now in
Table \ref{table_g_7_3_32} we have shown that 5 qubit Brown state
and 5 qubit Cluster state can be used for quantum dialogue if the
information are encoded by the elements of $G_{3}^{7}(32)$. This
is only a particular example, we have verified that dense coding can be done on  5 qubit Brown
state  in at least 6 different ways by using the
elements of $G_{3}^{1}(32),G_{3}^{2}(32),G_{3}^{4}(32),G_{3}^{5}(32),G_{3}^{7}(32),G_{3}^{8}(32)$
and similarly on  5 qubit Cluster state  in at least
4 different ways by using the elements of $G_{3}^{4}(32),G_{3}^{5}(32),G_{3}^{7}(32),G_{3}^{8}(32)$.
Thus quantum dialogue can be implemented in several ways by using
5 qubit states. In brief, our generalized protocol can be used to
implement quantum dialogue using Bell state, 4 and 5 qubit Cluster
states, $\Omega$ state, $Q_{4}$ state, $Q_{5}$ state, $W$ state,
$GHZ$ state, $GHZ$-\emph{like} state, 5 qubit Brown state etc.


%
\begin{table*}
\caption{\label{table:densecoding_GHZlikeandW_g_9_2_8}Dense coding of $GHZ$-\emph{like} states and $W$ states using the
elements of $G_{2}^{9}(8).$}
\begin{ruledtabular}
\begin{tabular}{ccc}

Unitary operations  & $|\lambda\rangle_{GHZ-like}=\frac{1}{\sqrt{2}}(|010\rangle+|100\rangle+$ & $|W\rangle_{0}=\frac{1}{2}(|0001\rangle+|0010\rangle+$\tabularnewline
on qubits 1 and 2 &$|001\rangle+|111\rangle)$ &$|0100\rangle+|1000\rangle)$\tabularnewline
\hline
$U_{0}=I\otimes I$ & $\frac{1}{\sqrt{2}}(|010\rangle+|100\rangle+|001\rangle+|111\rangle)$ & $\frac{1}{2}(|0001\rangle+|0010\rangle+|0100\rangle+|1000\rangle)$\tabularnewline

$U_{1}=Z\otimes Z$ & $\frac{1}{\sqrt{2}}(-|010\rangle-|100\rangle+|001\rangle+|111\rangle)$ & $\frac{1}{2}(|0001\rangle+|0010\rangle-|1000\rangle-|0100\rangle)$\tabularnewline

$U_{2}=X\otimes iY$ & $\frac{1}{\sqrt{2}}(|100\rangle-|010\rangle-|111\rangle+|001\rangle)$ & $\frac{1}{2}(-|1101\rangle-|1110\rangle+|1000\rangle-|0100\rangle)$\tabularnewline

$U_{3}=iY\otimes X$ & $\frac{1}{\sqrt{2}}(-|100\rangle+|010\rangle-|111\rangle+|001\rangle)$ & $\frac{1}{2}(-|1101\rangle-|1110\rangle-|1000\rangle+|0100\rangle)$\tabularnewline

$U_{4}=X\otimes I$ & $\frac{1}{\sqrt{2}}(|110\rangle+|000\rangle+|101\rangle+|011\rangle)$ & $\frac{1}{2}(|1001\rangle+|1010\rangle+|1100\rangle+|0000\rangle)$\tabularnewline

$U_{5}=iY\otimes Z$ & $\frac{1}{\sqrt{2}}(|110\rangle+|000\rangle-|101\rangle-|011\rangle)$ & $\frac{1}{2}(-|1001\rangle-|1010\rangle+|1100\rangle+|0000\rangle)$\tabularnewline

$U_{6}=Z\otimes X$ & $\frac{1}{\sqrt{2}}(|000\rangle-|110\rangle+|011\rangle-|101\rangle)$ & $\frac{1}{2}(|0101\rangle+|0110\rangle+|0000\rangle-|1100\rangle)$\tabularnewline

$U_{7}=I\otimes iY$ & $\frac{1}{\sqrt{2}}(|000\rangle-|110\rangle-|011\rangle+|101\rangle)$ & $\frac{1}{2}(-|0101\rangle-|0110\rangle+|0000\rangle-|1100\rangle)$\tabularnewline

\end{tabular}
\end{ruledtabular}
\end{table*}
\begin{table}
\caption{\label{table_g_8_2_8}Dense coding of $W$ states using the elements of $G_{2}^{8}(8).$}

\begin{ruledtabular}
\begin{tabular}{cc}

Unitary operations   & $|W\rangle_{0}=\frac{1}{2}(|0001\rangle+|0010\rangle+$\tabularnewline

on qubits 1 and 2& $|0100\rangle+|1000\rangle)$
\tabularnewline \hline
$U_{0}=I\otimes I$ & $\frac{1}{2}(|0001\rangle+|0010\rangle+|0100\rangle+|1000\rangle)$\tabularnewline

$U_{1}=Z\otimes Z$ & $\frac{1}{2}(|0001\rangle+|0010\rangle-|1000\rangle-|0100\rangle)$\tabularnewline

$U_{2}=X\otimes iY$ & $\frac{1}{2}(-|1101\rangle-|1110\rangle+|1000\rangle-|0100\rangle)$\tabularnewline

$U_{3}=iY\otimes X$ & $\frac{1}{2}(-|1101\rangle-|1110\rangle-|1000\rangle+|0100\rangle)$\tabularnewline

$U_{4}=I\otimes X$ & $\frac{1}{2}(|0101\rangle+|0110\rangle+|0000\rangle+|1100\rangle)$\tabularnewline

$U_{5}=Z\otimes iY$ & $\frac{1}{2}(-|0101\rangle-|0110\rangle+|0000\rangle+|1100\rangle)$\tabularnewline

$U_{6}=iY\otimes I$ & $\frac{1}{2}(-|1001\rangle-|1010\rangle-|1100\rangle+|0000\rangle)$\tabularnewline

$U_{7}=X\otimes Z$ & $\frac{1}{2}(|1001\rangle+|1010\rangle-|1100\rangle+|0000\rangle)$\tabularnewline

\end{tabular}
\end{ruledtabular}
\end{table}

\section{Conclusions}  \label{conclusion}

We have explored the underlying symmetry of the existing quantum dialogue
protocols and have shown that the information splitting plays a crucial
role in their implementation. Then we have obtained a sufficiency
condition for implementation of quantum dialogue. The sufficiency
condition obtained may be briefly stated as: If we have a set of mutually
orthogonal n-qubit states $\{|\phi_{0}\rangle,|\phi_{1}\rangle,....,|\phi_{i}\rangle,...,|\phi_{2^{n}-1}\rangle\}$
and a set of $m-qubit$ ($m\leq n$) unitary operators $\{U_{0},U_{1},U_{2},...,U_{2^{n}-1}\}$
such that the $U_{i}|\phi_{0}\rangle=|\phi_{i}\rangle$ and $\{U_{0},U_{1},U_{2},...,U_{2^{n}-1}\}$
forms a group under multiplication then it would be sufficient to
construct a quantum dialogue protocol of Ba An type using this set
of quantum states and this group of unitary operators. We have used
this sufficiency condition to provide a generalized algorithm for
implementation of quantum dialogue. All the existing Ba An type of
quantum dialogue protocol thus automatically become special cases
of our protocol. We have also provided a systematic way of generating
groups of unitary operators that are useful for implementation of
quantum dialogue and have shown that those groups may be used to implement
our generalized quantum dialogue protocol using Bell state, 4 qubit
and 5 qubit Cluster states, $\Omega$ state, $Q_{4}$ state, $Q_{5}$
state, $W$ state, $GHZ$ state, $GHZ$-\emph{like} state, Brown state
etc. To reach at this conclusion, we have used several Tables of dense
coding. Here we have obtained them through a systematic and independent
approach, where all the unitary operators of a group or subgroup of
appropriate order are applied on the quantum state of interest and
the orthogonality of the outputs are checked. To be precise, if we
wish to study the possibility of implementing quantum dialogue using
3-qubit $GHZ$ states then we would apply all the order 8 subgroups
of $G_{2}$ on the $GHZ$ state and see which subgroups yield a set
of 8 mutually orthogonal states. For example, we found that $G_{2}^{1}(8),\, G_{2}^{2}(8),\, G_{2}^{4}(8)$
and $G_{2}^{5}(8)$ are such subgroups which provides us \emph{useful
dense coding} %
\footnote{By useful dense coding we mean that the operators used for dense coding
forms a group and consequently, the combination of the group of unitary
operators and the quantum state is useful for quantum dialogue.%
} scheme. Our systematic approach has produced a large number of \emph{useful
dense coding} schemes. A few of these dense coding schemes can also
be found in the existing literature. For example, \emph{useful dense
coding} operations for 4 qubit Cluster state is discussed in \cite{Tsai-cluster}
in a different context (in context of DSQC). Similarly, in a different
context \emph{useful dense coding} schemes (only one scheme for each
case) for the $\Omega$ state, $Q_{4}$ state, $Q_{5}$ state, $W$
state are reported in \cite{Pradhan-Agarwal-Pati}, also in \cite{Panigrahi-PRA}
a dense coding scheme for 5 qubit Brown states is reported%
\footnote{In Table V of \cite{Panigrahi-PRA} dense coding of 5 qubit Brown state
is attempted. The table has some typos. Unitary operations of the
corrected table will coincide with $G_{3}^{1}(32).$%
}.
Our systematic approach have yielded additional schemes of\emph{
useful dense coding }using $GHZ$, $W$, $Q_{4},$ $Q_{5}$ and Brown
states and new examples of \emph{useful dense coding} schemes using
$GHZ-$\emph{like} states and 5 qubit Cluster states. The present
paper is focused on quantum dialogue but the newly found \emph{useful
dense coding} schemes can find their applications on other aspects
of quantum communication (e.g. DSQC) and quantum games. It is not
our purpose to discuss those possibilities here. Rather we would like
to note that our group theoretic approach provides a wide choice of
possible quantum states that can be used for quantum dialogue.

\begin{table}
\caption{\label{table_g_3_2_8}Dense coding of $GHZ$-\emph{like} state by using the elements of
$G_{2}^{3}(8)$. }
\begin{ruledtabular}
\begin{tabular}{cc}

Unitary operations &  $|\lambda\rangle_{GHZ-like}=\frac{1}{\sqrt{2}}(|010\rangle+|100\rangle+$\tabularnewline
on qubits 1 and 2& $|001\rangle+|111\rangle)$\tabularnewline\hline
$U_{0}=I\otimes I$ & $\frac{1}{\sqrt{2}}(|010\rangle+|100\rangle+|001\rangle+|111\rangle)$\tabularnewline

$U_{1}=Z\otimes I$ & $\frac{1}{\sqrt{2}}(|010\rangle-|100\rangle+|001\rangle-|111\rangle)$\tabularnewline

$U_{2}=X\otimes I$ & $\frac{1}{\sqrt{2}}(|110\rangle+|000\rangle+|101\rangle+|011\rangle)$\tabularnewline

$U_{3}=iY\otimes I$ & $\frac{1}{\sqrt{2}}(-|110\rangle+|000\rangle-|101\rangle+|011\rangle)$\tabularnewline

$U_{4}=I\otimes Z$ & $\frac{1}{\sqrt{2}}(-|010\rangle+|100\rangle+|001\rangle-|111\rangle)$\tabularnewline

$U_{5}=Z\otimes Z$ & $\frac{1}{\sqrt{2}}(-|010\rangle-|100\rangle+|001\rangle+|111\rangle)$\tabularnewline

$U_{6}=X\otimes Z$ & $\frac{1}{\sqrt{2}}(-|110\rangle+|000\rangle+|101\rangle-|011\rangle)$\tabularnewline

$U_{7}=iY\otimes Z$ & $\frac{1}{\sqrt{2}}(|110\rangle+|000\rangle-|101\rangle-|011\rangle)$\tabularnewline

\end{tabular}
\end{ruledtabular}
\end{table}

\begin{table*}
\caption{\label{table_g_6_2_8}Dense coding of $GHZ$\emph{-like} and $Q_{4}$ states by using the
elements of $G_{2}^{6}(8)$. }
\begin{ruledtabular}
\begin{tabular}{ccc}

Unitary Operations  & $|\lambda\rangle_{GHZ-like}=\frac{1}{\sqrt{2}}(|010\rangle+|100\rangle+|001\rangle+|111\rangle)$ & $|Q_{4}\rangle_{0}=\frac{1}{2}(|0000\rangle+|0101\rangle+|1000\rangle+|1110\rangle)$\tabularnewline
on qubits 1 and 2 &  & \tabularnewline
\hline
$U_{0}=I\otimes I$ & $\frac{1}{\sqrt{2}}(|010\rangle+|100\rangle+|001\rangle+|111\rangle)$ & $\frac{1}{2}(|0000\rangle+|0101\rangle+|1000\rangle+|1110\rangle)$\tabularnewline

$U_{1}=I\otimes Z$ & $\frac{1}{\sqrt{2}}(-|010\rangle+|100\rangle+|001\rangle-|111\rangle)$ & $\frac{1}{2}(|0000\rangle-|0101\rangle+|1000\rangle-|1110\rangle)$\tabularnewline

$U_{2}=I\otimes X$ & $\frac{1}{\sqrt{2}}(|000\rangle+|110\rangle+|011\rangle+|101\rangle)$ & $\frac{1}{2}(|0100\rangle+|0001\rangle+|1100\rangle+|1010\rangle)$\tabularnewline

$U_{3}=I\otimes iY$ & $\frac{1}{\sqrt{2}}(|000\rangle-|110\rangle-|011\rangle+|101\rangle)$ & $\frac{1}{2}(-|0100\rangle+|0001\rangle-|1100\rangle+|1010\rangle)$\tabularnewline

$U_{4}=Z\otimes I$ & $\frac{1}{\sqrt{2}}(|010\rangle-|100\rangle+|001\rangle-|111\rangle)$ & $\frac{1}{2}(|0000\rangle+|0101\rangle-|1000\rangle-|1110\rangle)$\tabularnewline

$U_{5}=Z\otimes Z$ & $\frac{1}{\sqrt{2}}(-|010\rangle-|100\rangle+|001\rangle+|111\rangle)$ & $\frac{1}{2}(|0000\rangle-|0101\rangle-|1000\rangle+|1110\rangle)$\tabularnewline

$U_{6}=Z\otimes X$ & $\frac{1}{\sqrt{2}}(|000\rangle-|110\rangle+|011\rangle-|101\rangle)$ & $\frac{1}{2}(|0100\rangle+|0001\rangle-|1100\rangle-|1010\rangle)$\tabularnewline

$U_{7}=Z\otimes iY$ & $\frac{1}{\sqrt{2}}(|000\rangle+|110\rangle-|011\rangle-|101\rangle)$ & $\frac{1}{2}(-|0100\rangle+|0001\rangle+|1100\rangle-|1010\rangle)$\tabularnewline

\end{tabular}
\end{ruledtabular}
\end{table*}

\begin{table}
\caption{\label{table_g_7_2_8}Dense coding of $Q_{4}$ states using the elements of $G_{2}^{7}(8).$ The operators are applied on qubits 1 and 2.}
\begin{ruledtabular}
\begin{tabular}{ccc}

Unitary  & Operators  & $|Q_{4}\rangle_{0}=\frac{1}{2}(|0000\rangle+|0101\rangle+$\tabularnewline
operations &  & $|1000\rangle+|1110\rangle)$\tabularnewline\hline
$U_{0}$ & $I\otimes I$ & $\frac{1}{2}(|0000\rangle+|0101\rangle+|1000\rangle+|1110\rangle)$\tabularnewline

$U_{1}$ & $I\otimes Z$ & $\frac{1}{2}(|0000\rangle-|0101\rangle+|1000\rangle-|1110\rangle)$\tabularnewline

$U_{2}$ & $Z\otimes I$ & $\frac{1}{2}(|0000\rangle+|0101\rangle-|1000\rangle-|1110\rangle)$\tabularnewline

$U_{3}$ & $Z\otimes Z$ & $\frac{1}{2}(|0000\rangle-|0101\rangle-|1000\rangle+|1110\rangle)$\tabularnewline

$U_{4}$ & $X\otimes X$ & $\frac{1}{2}(|1100\rangle+|1001\rangle+|0100\rangle+|0010\rangle)$\tabularnewline

$U_{5}$ & $iY\otimes X$ & $\frac{1}{2}(-|1100\rangle-|1001\rangle+|0100\rangle+|0010\rangle)$\tabularnewline

$U_{6}$ & $X\otimes iY$ & $\frac{1}{2}(-|1100\rangle+|1001\rangle-|0100\rangle+|0010\rangle)$\tabularnewline

$U_{7}$ & $iY\otimes iY$ & $\frac{1}{2}(|1100\rangle-|1001\rangle-|0100\rangle+|0010\rangle)$\tabularnewline

\end{tabular}
\end{ruledtabular}
\end{table}

\newpage
\begin{table}
\caption{\label{table_g_4_2_8}Dense coding of $Q_{5}$ states using the elements of $G_{2}^{4}(8).$ The operators are applied on qubits 1 and 2.}
\begin{ruledtabular}
\begin{tabular}{ccc}

Unitary  & Operators   & $|Q_{5}\rangle_{0}=\frac{1}{2}(|0000\rangle+|1011\rangle+$\tabularnewline
operations &  & $|1101\rangle+|1110\rangle)$ \tabularnewline\hline
$U_{0}$ & $I\otimes I$ & $\frac{1}{2}(|0000\rangle+|1011\rangle+|1101\rangle+|1110\rangle)$\tabularnewline

$U_{1}$ & $X\otimes I$ & $\frac{1}{2}(|1000\rangle+|0011\rangle+|0101\rangle+|0110\rangle)$\tabularnewline

$U_{2}$ & $I\otimes X$ & $\frac{1}{2}(|0100\rangle+|1111\rangle+|1001\rangle+|1010\rangle)$\tabularnewline

$U_{3}$ & $X\otimes X$ & $\frac{1}{2}(|1100\rangle+|0111\rangle+|0001\rangle+|0010\rangle)$\tabularnewline

$U_{4}$ & $I\otimes iY$ & $\frac{1}{2}(-|0100\rangle-|1111\rangle+|1001\rangle+|1010\rangle)$\tabularnewline

$U_{5}$ & $X\otimes iY$ & $\frac{1}{2}(-|1100\rangle-|0111\rangle+|0001\rangle+|0010\rangle)$\tabularnewline

$U_{6}$ & $I\otimes Z$ & $\frac{1}{2}(|0000\rangle+|1011\rangle-|1101\rangle-|1110\rangle)$\tabularnewline

$U_{7}$ & $X\otimes Z$ & $\frac{1}{2}(|1000\rangle+|0011\rangle-|0101\rangle-|0110\rangle)$\tabularnewline

\end{tabular}
\end{ruledtabular}
\end{table}

Quantum dialogue can be implemented by using same quantum state but
by using different groups of unitary operators as discussed above
and summarized in Table \ref{table_summary}. For example, we have
shown that there exist at least 4 different groups of unitary operators
($G_{2}^{1}(8),G_{2}^{2}(8),G_{2}^{4}(8)$ and $G_{2}^{5}(8)$) that
may be used to implement quantum dialogue using $GHZ$ state. Similarly
there exist at least six different ways (by using $G_{2}^{2}(8),G_{2}^{3}(8),G_{2}^{5}(8),G_{2}^{6}(8),G_{2}^{8}(8)$
and $G_{2}^{9}(8)$) to implement quantum dialogue using $GHZ$-\emph{like}
states, six different ways to implement quantum dialogue using 5 qubit
Brown state (by using $G_{3}^{1}(32),G_{3}^{2}(32),G_{3}^{4}(32),G_{3}^{5}(32),G_{3}^{7}(32)$
and $G_{3}^{8}(32)$), four different ways (by using $G_{3}^{4}(32),G_{3}^{5}(32),G_{3}^{7}(32),G_{3}^{8}(32)$ ) to implement
quantum dialogue using 5 qubit Cluster state, three different ways
(by using $G_{2}^{3}(8),G_{2}^{4}(8)$ and $G_{2}^{5}(8)$ ) to implement
quantum dialogue using $Q_{5}$ state and so on. If we closely look
at these groups and subgroups of unitary operators we will observe
an intrinsic symmetry. For example, $G_{3}^{1}(32)=G_{2}\otimes\{I,X\},G_{3}^{4}(32)=\{I,X\}\otimes G_{2}$
and $G_{3}^{7}(32)=G_{1}\otimes\{I,X\}\otimes G_{1}$ are similar only
the positions of specific Pauli operators are changing or in other
words they are equivalent under swapping of qubits on which a particular
set of operators are applied. Thus once we find out a particular group
of unitary operators for dense coding of a quantum state then from
the permutation symmetry of the quantum state we can conclude whether
other similar groups will do the dense coding or not.

Our results can  be easily extended to secured multi-party computation (SMC) tasks. As an example, here we will briefly describe how our results can be used  to obtain solution of a specific SMC task which is known as socialist millionaire problem \cite{smp}. In this problem two millionaire wish to compare their wealth but they do not want to disclose the amount of their wealth to each other. This problem is also referred as the problem of private comparison of equal information. Now if we consider that  Alice and Bob are the millionaire who wants to compare their wealth and Charlie is a semi-honest third party then our protocol works as follows: Charlie prepares an n-qubit entangled state in one of the possible mutually orthogonal states  $\{|\phi_{0}\rangle,|\phi_{1}\rangle,....,|\phi_{i}\rangle,...,|\phi_{2^{n}-1}\rangle\}$ (say he prepares it in $|\phi_{i}\rangle$) and keeps the home photons with himself and sends the travel photons to Alice, who encodes her information (the value of her wealth) by applying unitary operations $\{U_{0},U_{1},U_{2},...,U_{2^{n}-1}\}$ as per the encoding rule. Then Alice sends the encoded qubits to Bob. As Bob has access to only travel photons and since he does not know initial state he can not obtain the information encoded by Alice. Now Bob also encodes his information (the value of his wealth) by using the same set of encoding operations $\{U_{0},U_{1},U_{2},...,U_{2^{n}-1}\}$  and sends the qubits to Charlie. Now if  $\{U_{0},U_{1},U_{2},...,U_{2^{n}-1}\}$ forms a group under multiplication then Charlie will obtain one of the mutually orthogonal state.  Charlie can measure the final state using  $\{|\phi_{0}\rangle,|\phi_{1}\rangle,....,|\phi_{i}\rangle,...,|\phi_{2^{n}-1}\rangle\}$ as basis and deterministically obtain the final state $|\phi_{f}\rangle$. Till this point this protocol  is similar to the quantum dialogue protocol. The difference between the two protocols is that instead of Alice now Charlie prepares the initial state and Charlie knows nothing about Alice and Bob's encoding. He knows only the final state and initial state, so his knowledge is same as that of  Eve in the previous protocol.  If Charlie finds that the initial state and the final state are same  (i.e. $|\phi_{i}\rangle=|\phi_{f}\rangle$) then  the classical information encoded by  Alice and Bob are same. In all other cases the classical information encoded by Alice and Bob are different.  If we consider the encoded information as the amount of  their assets then it solves socialist millionaire problem. Neither Alice, nor Bob, nor even Charlie knows how much assets are there in possession of other. Since classical broadcasting is not required here, the intrinsic problem of information leakage in quantum dialogue protocol is not present here. Thus the wide range of unitary operators and quantum states obtained here for implementation of quantum dialogue  protocol (as summarized in  Table \ref{table_summary}) can  also be used to provide solutions of socialist millionaire problem.

\newpage
Here it would be apt to note that there also exist a hierarchy among
quantum communication protocols. For example, any quantum dialogue
protocol may be reduced to a DSQC protocol by considering that Bob
does not send any meaningful information (does not encode any information),
rather he just randomly sends one of the state $\{|\phi_{0}\rangle,|\phi_{1}\rangle,....,|\phi_{i}\rangle,...,|\phi_{2^{n}-1}\rangle\}$
to Alice and Alice encodes secret information as per the above protocol.
Now it is well known that all DSQC and QSDC protocols can be reduced
to protocols of QKD. It is easy to visualize, if instead of a meaningful
information Alice sends a random key to Bob then a DSQC or QSDC protocol
will become a QKD protocol. Similarly in a quantum dialogue protocol
if Alice and Bob both sends each other a set of random bits then at
the end of successful conversation they will be able to generate two
sets of unconditionally secure key. Thus a quantum dialogue protocol
can always be used to implement DSQC and QKD protocol but the converse
is not true.
\newline
From the perspective of experimental implementation it is a very attractive
situation as there are so many options to implement the same task
(quantum dialogue and solution of socialist millionaire problem) and thus the current work is expected to motivate
experimentalist. It is indeed attractive but the situation is actually
more favorable because we have just given a handful of examples. A
simple computer program can generate all the relevant subgroups of
$G_{n}$ and the set of states that are unitarily connected by the
elements of a particular subgroup or group.

%
\begin{table*}
\caption{\label{table_g_7_3_32}Dense coding of 5 qubit Brown state and Cluster state using the elements
of $G_{3}^{7}(32).$}
\begin{ruledtabular}
\begin{tabular}{ccc}

Unitary operators  & $|C_{5}\rangle=\frac{1}{2}[|00000\rangle+|00111\rangle+|11101\rangle+|11010\rangle]$ & $|\psi\rangle_{Brown}=\frac{1}{2}[|001\rangle|\phi_{-}\rangle+|010\rangle|\psi_{-}\rangle+|100\rangle|\phi_{+}\rangle+|111\rangle|\psi_{+}\rangle]$\tabularnewline
on 1, 2 and 3 qubits & & \tabularnewline \hline
$U_{0}=I\otimes I\otimes I$ & $\frac{1}{2}[+|00000\rangle+|00111\rangle+|11101\rangle+|11010\rangle]$ & $\frac{1}{2}[+|001\rangle|\phi_{-}\rangle+|010\rangle|\psi_{-}\rangle+|100\rangle|\phi_{+}\rangle+|111\rangle|\psi_{+}\rangle]$\tabularnewline

$U_{1}=X\otimes I\otimes I$ & $\frac{1}{2}[+|10000\rangle+|10111\rangle+|01101\rangle+|01010\rangle]$ & $\frac{1}{2}[+|101\rangle|\phi_{-}\rangle+|110\rangle|\psi_{-}\rangle+|000\rangle|\phi_{+}\rangle+|011\rangle|\psi_{+}\rangle]$\tabularnewline

$U_{2}=iY\otimes I\otimes I$ & $\frac{1}{2}[-|10000\rangle-|10111\rangle+|01101\rangle+|01010\rangle]$ & $\frac{1}{2}[-|101\rangle|\phi_{-}\rangle-|110\rangle|\psi_{-}\rangle+|000\rangle|\phi_{+}\rangle+|011\rangle|\psi_{+}\rangle]$\tabularnewline

$U_{3}=Z\otimes I\otimes I$ & $\frac{1}{2}[+|00000\rangle+|00111\rangle-|11101\rangle-|11010\rangle]$ & $\frac{1}{2}[+|001\rangle|\phi_{-}\rangle+|010\rangle|\psi_{-}\rangle-|100\rangle|\phi_{+}\rangle-|111\rangle|\psi_{+}\rangle]$\tabularnewline

$U_{4}=I\otimes I\otimes X$ & $\frac{1}{2}[+|00100\rangle+|00011\rangle+|11001\rangle+|11110\rangle]$ & $\frac{1}{2}[+|000\rangle|\phi_{-}\rangle+|011\rangle|\psi_{-}\rangle+|101\rangle|\phi_{+}\rangle+|110\rangle|\psi_{+}\rangle]$\tabularnewline

$U_{5}=X\otimes I\otimes X$ & $\frac{1}{2}[+|10100\rangle+|10011\rangle+|01001\rangle+|01110\rangle]$ & $\frac{1}{2}[+|100\rangle|\phi_{-}\rangle+|111\rangle|\psi_{-}\rangle+|001\rangle|\phi_{+}\rangle+|010\rangle|\psi_{+}\rangle]$\tabularnewline

$U_{6}=iY\otimes I\otimes X$ & $\frac{1}{2}[-|10100\rangle-|10011\rangle+|01001\rangle+|01110\rangle]$ & $\frac{1}{2}[-|100\rangle|\phi_{-}\rangle-|111\rangle|\psi_{-}\rangle+|001\rangle|\phi_{+}\rangle+|010\rangle|\psi_{+}\rangle]$\tabularnewline

$U_{7}=Z\otimes I\otimes X$ & $\frac{1}{2}[+|00100\rangle+|00011\rangle-|11001\rangle-|11110\rangle]$ & $\frac{1}{2}[+|000\rangle|\phi_{-}\rangle+|011\rangle|\psi_{-}\rangle-|101\rangle|\phi_{+}\rangle-|110\rangle|\psi_{+}\rangle]$\tabularnewline
$U_{8}=I\otimes I\otimes iY$ & $\frac{1}{2}[-|00100\rangle+|00011\rangle+|11001\rangle-|11110\rangle]$ & $\frac{1}{2}[+|000\rangle|\phi_{-}\rangle-|011\rangle|\psi_{-}\rangle-|101\rangle|\phi_{+}\rangle+|110\rangle|\psi_{+}\rangle]$\tabularnewline

$U_{9}=X\otimes I\otimes iY$ & $\frac{1}{2}[-|10100\rangle+|10011\rangle+|01001\rangle-|01110\rangle]$ & $\frac{1}{2}[+|100\rangle|\phi_{-}\rangle-|111\rangle|\psi_{-}\rangle-|001\rangle|\phi_{+}\rangle+|010\rangle|\psi_{+}\rangle]$\tabularnewline

$U_{10}=iY\otimes I\otimes iY$ & $\frac{1}{2}[+|10100\rangle-|10011\rangle+|01001\rangle-|01110\rangle]$ & $\frac{1}{2}[-|100\rangle|\phi_{-}\rangle+|111\rangle|\psi_{-}\rangle-|001\rangle|\phi_{+}\rangle+|010\rangle|\psi_{+}\rangle]$\tabularnewline

$U_{11}=Z\otimes I\otimes iY$ & $\frac{1}{2}[-|00100\rangle+|00011\rangle-|11001\rangle+|11110\rangle]$ & $\frac{1}{2}[+|000\rangle|\phi_{-}\rangle-|011\rangle|\psi_{-}\rangle+|101\rangle|\phi_{+}\rangle-|110\rangle|\psi_{+}\rangle]$\tabularnewline

$U_{12}=I\otimes I\otimes Z$ & $\frac{1}{2}[+|00000\rangle-|00111\rangle-|11101\rangle+|11010\rangle]$ & $\frac{1}{2}[-|001\rangle|\phi_{-}\rangle+|010\rangle|\psi_{-}\rangle+|100\rangle|\phi_{+}\rangle-|111\rangle|\psi_{+}\rangle]$\tabularnewline

$U_{13}=X\otimes I\otimes Z$ & $\frac{1}{2}[+|10000\rangle-|10111\rangle-|01101\rangle+|01010\rangle]$ & $\frac{1}{2}[-|101\rangle|\phi_{-}\rangle+|110\rangle|\psi_{-}\rangle+|000\rangle|\phi_{+}\rangle-|011\rangle|\psi_{+}\rangle]$\tabularnewline

$U_{14}=iY\otimes I\otimes Z$ & $\frac{1}{2}[-|10000\rangle+|10111\rangle-|01101\rangle+|01010\rangle]$ & $\frac{1}{2}[+|101\rangle|\phi_{-}\rangle-|110\rangle|\psi_{-}\rangle+|000\rangle|\phi_{+}\rangle-|011\rangle|\psi_{+}\rangle]$\tabularnewline

$U_{15}=Z\otimes I\otimes Z$ & $\frac{1}{2}[+|00000\rangle-|00111\rangle+|11101\rangle-|11010\rangle]$ & $\frac{1}{2}[-|001\rangle|\phi_{-}\rangle+|010\rangle|\psi_{-}\rangle-|100\rangle|\phi_{+}\rangle+|111\rangle|\psi_{+}\rangle]$\tabularnewline

$U_{16}=I\otimes X\otimes I$ & $\frac{1}{2}[+|01000\rangle+|01111\rangle+|10101\rangle+|10010\rangle]$ & $\frac{1}{2}[+|011\rangle|\phi_{-}\rangle+|000\rangle|\psi_{-}\rangle+|110\rangle|\phi_{+}\rangle+|101\rangle|\psi_{+}\rangle]$\tabularnewline

$U_{17}=X\otimes X\otimes I$ & $\frac{1}{2}[+|11000\rangle+|11111\rangle+|00101\rangle+|00010\rangle]$ & $\frac{1}{2}[+|111\rangle|\phi_{-}\rangle+|100\rangle|\psi_{-}\rangle+|010\rangle|\phi_{+}\rangle+|001\rangle|\psi_{+}\rangle]$\tabularnewline

$U_{18}=iY\otimes X\otimes I$ & $\frac{1}{2}[-|11000\rangle-|11111\rangle+|00101\rangle+|00010\rangle]$ & $\frac{1}{2}[-|111\rangle|\phi_{-}\rangle-|100\rangle|\psi_{-}\rangle+|010\rangle|\phi_{+}\rangle+|001\rangle|\psi_{+}\rangle]$\tabularnewline

$U_{19}=Z\otimes X\otimes I$ & $\frac{1}{2}[+|01000\rangle+|01111\rangle-|10101\rangle-|10010\rangle]$ & $\frac{1}{2}[+|011\rangle|\phi_{-}\rangle+|000\rangle|\psi_{-}\rangle-|110\rangle|\phi_{+}\rangle-|101\rangle|\psi_{+}\rangle]$\tabularnewline

$U_{20}=I\otimes X\otimes X$ & $\frac{1}{2}[+|01100\rangle+|01011\rangle+|10001\rangle+|10110\rangle]$ & $\frac{1}{2}[+|010\rangle|\phi_{-}\rangle+|001\rangle|\psi_{-}\rangle+|111\rangle|\phi_{+}\rangle+|100\rangle|\psi_{+}\rangle]$\tabularnewline

$U_{21}=X\otimes X\otimes X$ & $\frac{1}{2}[+|11100\rangle+|11011\rangle+|00001\rangle+|00110\rangle]$ & $\frac{1}{2}[+|110\rangle|\phi_{-}\rangle+|101\rangle|\psi_{-}\rangle+|011\rangle|\phi_{+}\rangle+|000\rangle|\psi_{+}\rangle]$\tabularnewline

$U_{22}=iY\otimes X\otimes X$ & $\frac{1}{2}[-|11100\rangle-|11011\rangle+|00001\rangle+|00110\rangle]$ & $\frac{1}{2}[-|110\rangle|\phi_{-}\rangle-|101\rangle|\psi_{-}\rangle+|011\rangle|\phi_{+}\rangle+|000\rangle|\psi_{+}\rangle]$\tabularnewline

$U_{23}=Z\otimes X\otimes X$ & $\frac{1}{2}[+|01100\rangle+|01011\rangle-|10001\rangle-|10110\rangle]$ & $\frac{1}{2}[+|010\rangle|\phi_{-}\rangle+|001\rangle|\psi_{-}\rangle-|111\rangle|\phi_{+}\rangle-|100\rangle|\psi_{+}\rangle]$\tabularnewline

$U_{24}=I\otimes X\otimes iY$ & $\frac{1}{2}[-|01100\rangle+|01011\rangle+|10001\rangle-|10110\rangle]$ & $\frac{1}{2}[+|010\rangle|\phi_{-}\rangle-|001\rangle|\psi_{-}\rangle-|111\rangle|\phi_{+}\rangle+|100\rangle|\psi_{+}\rangle]$\tabularnewline

$U_{25}=X\otimes X\otimes iY$ & $\frac{1}{2}[-|11100\rangle+|11011\rangle+|00001\rangle-|00110\rangle]$ & $\frac{1}{2}[+|110\rangle|\phi_{-}\rangle-|101\rangle|\psi_{-}\rangle-|011\rangle|\phi_{+}\rangle+|000\rangle|\psi_{+}\rangle]$\tabularnewline

$U_{26}=iY\otimes X\otimes iY$ & $\frac{1}{2}[+|11100\rangle-|11011\rangle+|00001\rangle-|00110\rangle]$ & $\frac{1}{2}[-|110\rangle|\phi_{-}\rangle+|101\rangle|\psi_{-}\rangle-|011\rangle|\phi_{+}\rangle+|000\rangle|\psi_{+}\rangle]$\tabularnewline

$U_{27}=Z\otimes X\otimes iY$ & $\frac{1}{2}[-|01100\rangle+|01011\rangle-|10001\rangle+|10110\rangle]$ & $\frac{1}{2}[+|010\rangle|\phi_{-}\rangle-|001\rangle|\psi_{-}\rangle+|111\rangle|\phi_{+}\rangle-|100\rangle|\psi_{+}\rangle]$\tabularnewline

$U_{28}=I\otimes X\otimes Z$ & $\frac{1}{2}[+|01000\rangle-|01111\rangle-|10101\rangle+|10010\rangle]$ & $\frac{1}{2}[-|011\rangle|\phi_{-}\rangle+|000\rangle|\psi_{-}\rangle+|110\rangle|\phi_{+}\rangle-|101\rangle|\psi_{+}\rangle]$\tabularnewline

$U_{29}=X\otimes X\otimes Z$ & $\frac{1}{2}[+|11000\rangle-|11111\rangle-|00101\rangle+|00010\rangle]$ & $\frac{1}{2}[-|111\rangle|\phi_{-}\rangle+|100\rangle|\psi_{-}\rangle+|010\rangle|\phi_{+}\rangle-|001\rangle|\psi_{+}\rangle]$\tabularnewline

$U_{30}=iY\otimes X\otimes Z$ & $\frac{1}{2}[-|11000\rangle+|11111\rangle-|00101\rangle+|00010\rangle]$ & $\frac{1}{2}[+|111\rangle|\phi_{-}\rangle-|100\rangle|\psi_{-}\rangle+|010\rangle|\phi_{+}\rangle-|001\rangle|\psi_{+}\rangle]$\tabularnewline

$U_{31}=Z\otimes X\otimes Z$ & $\frac{1}{2}[+|01000\rangle-|01111\rangle+|10101\rangle-|10010\rangle]$ & $\frac{1}{2}[-|011\rangle|\phi_{-}\rangle+|000\rangle|\psi_{-}\rangle-|110\rangle|\phi_{+}\rangle+|101\rangle|\psi_{+}\rangle]$\tabularnewline

\end{tabular}
\end{ruledtabular}
\end{table*}

\begin{table*}
\caption{\label{table_summary}List of useful quantum states and corresponding operators that may
be used to implement our generalized protocol of quantum dialogue. }

\begin{ruledtabular}
\begin{tabular}{cc}

Quantum state &  Group of unitary operations \tabularnewline
\hline
4 qubit $Q_{4}$ & $G_{2}^{6}(8),G_{2}^{7}(8)$\tabularnewline

3 qubit $GHZ$ & $G_{2}^{1}(8),G_{2}^{2}(8),G_{2}^{4}(8),G_{2}^{5}(8)$\tabularnewline

3 qubit $GHZ$-\emph{like} & $G_{2}^{2}(8),G_{2}^{3}(8),G_{2}^{5}(8),G_{2}^{6}(8),G_{2}^{8}(8),G_{2}^{9}(8)$\tabularnewline

4 qubit $W$ & $G_{2}^{8}(8),G_{2}^{9}(8)$\tabularnewline

4 qubit $Q_{5}$ & $G_{2}^{3}(8),G_{2}^{4}(8),G_{2}^{5}(8)$\tabularnewline

4 qubit Cluster state & $G_{2}$\tabularnewline

4 qubit $\Omega$ state & $G_{2}$\tabularnewline

2 qubit Bell state & $G_{1}$\tabularnewline

5 qubit Brown state & $G_{3}^{1}(32),G_{3}^{2}(32),G_{3}^{4}(32),G_{3}^{5}(32),G_{3}^{7}(32),G_{3}^{8}(32)$\tabularnewline

5 qubit Cluster state & $G_{3}^{4}(32),G_{3}^{5}(32),G_{3}^{7}(32),G_{3}^{8}(32)$\tabularnewline

\end{tabular}
\end{ruledtabular}
\end{table*}

\pagebreak

~
~

~
~

~
~
~
~

~
~

~
~
\newpage


\begin{acknowledgments}AP thanks Department of Science and Technology
(DST), India for support provided through the DST project No. SR/S2/LOP/2010.
AP also thanks the Ministry of Education of the Czech Republic for
support provided through the project CZ.1.05/2.1.00/03.0058.\end{acknowledgments}


\end{document}